\documentclass[12pt, preprint]{aastex}

\usepackage{natbib}

\newcommand{\Om}{\Omega_{\rm M}}
\newcommand{\Ol}{\Omega_{\Lambda}}
\newcommand{\Ob}{\Omega_b}

\newcommand{\Lbox}{L_{\rm box}}

\newcommand{\lam}{\lambda}
\newcommand{\Lam}{\Lambda}

\newcommand{\Del}{\Delta}
\newcommand{\hinv}{{h^{-1}}}
\newcommand{\mpc}{{\rm\,Mpc}}
\newcommand{\kpc}{{\rm\,kpc}}
\newcommand{\himpc}{\hinv{\rm\,Mpc}}
\newcommand{\hikpc}{\hinv{\rm\,kpc}}

\newcommand{\Gyr}{{\rm\,Gyr}}
\newcommand{\yr}{{\rm yr}}
\newcommand{\Msun}{{\rm M}_{\odot}}
\newcommand{\himsun}{\hinv{\Msun}}

\newcommand{\gtsim}{\gtrsim}
\newcommand{\Mstar}{M_{\rm star}}
\newcommand{\dd}{{\rm d}}

\slugcomment{DRAFT VERSION}

\shorttitle{Massive Galaxies at $z=2$}
\shortauthors{Nagamine et al.}

\begin{document}

\title{Massive galaxies in cosmological simulations: UV-selected sample at
 redshift $z=2$}

\author{Kentaro Nagamine\altaffilmark{1}, Renyue Cen\altaffilmark{2}, Lars Hernquist\altaffilmark{1}, Jeremiah P. Ostriker\altaffilmark{2,3},\\ \& Volker Springel\altaffilmark{4}}


\altaffiltext{1}{Harvard-Smithsonian Center for Astrophysics, 
60 Garden Street, Cambridge, MA 02138, U.S.A. \\
Email: knagamin@cfa.harvard.edu}

\altaffiltext{2}{Princeton University Observatory, Princeton, NJ 08544, U.S.A.}

\altaffiltext{3}{Institute of Astronomy, University of Cambridge, Madingley Road, 
Cambridge, CB3, OHA, UK}

\altaffiltext{4}{Max-Planck-Institut f\"{u}r Astrophysik, 
        Karl-Schwarzschild-Stra\ss{}e 1, 85740 Garching bei 
        M\"{u}nchen, Germany}


\begin{abstract}
We study the properties of galaxies at redshift $z=2$ in a $\Lam$ cold
dark matter ($\Lam$CDM) universe, using two different types of
hydrodynamic simulation methods -- Eulerian TVD and smoothed particle
hydrodynamics (SPH) -- and a
spectrophotometric analysis in the $U_n, G, R$ filter set.  The simulated
galaxies at $z=2$ satisfy the color-selection criteria proposed by
\citet{Ade04} and \citet{Steidel04} when we assume Calzetti extinction
with $E(B-V)=0.15$.  We find that the number density of simulated
galaxies brighter than $R<25.5$ at $z=2$ is about $2\times
10^{-2}~h^3~\mpc^{-3}$ for $E(B-V)=0.15$ in our most representative run, 
roughly one order of magnitude larger than that of Lyman break 
galaxies at $z=3$.  The most massive galaxies at $z=2$
have stellar masses $\gtsim 10^{11}\Msun$, and their observed-frame $G-R$
colors lie in the range $0.0<G-R<1.0$. They typically have been
continuously forming stars with a rate exceeding $30~\Msun~\yr^{-1}$ over
a few Gyrs from $z=10$ to $z=2$, although the TVD simulation indicates a
more sporadic star formation history than the SPH simulations.  Of order
half of their stellar mass was already assembled by $z\sim 4$.  The
bluest galaxies with colors $-0.2<G-R<0.0$ at $z=2$ are somewhat less
massive, with $\Mstar<10^{11}\himsun$, and lack a prominent old stellar
population.  On the other hand, the reddest massive galaxies at $z=2$
with $G-R \geq 1.0$ and $\Mstar>10^{10}\himsun$ completed the build-up of
their stellar mass by $z\sim 3$.  Interestingly, our study indicates that
the majority of the most massive galaxies at $z=2$ should be detectable
at rest-frame ultra-violet wavelengths, contrary to some recent claims
made on the basis of near-infrared studies of galaxies at the same epoch,
provided the median extinction is less than $E(B-V)<0.3$ as indicated by
surveys of Lyman break galaxies at $z=3$.  However, our results also
suggest that the fraction of stellar mass contained in galaxies that pass
the color-selection criteria used by \citet{Steidel04} could be as low as
50\% of the total stellar mass in the Universe at $z=2$. Our simulations
imply that the missing stellar mass is contained in fainter ($R>25.5$) 
and intrinsically redder galaxies. 
The bright end of the rest-frame $V$-band luminosity function of $z=2$ 
galaxies can be characterized by a Schechter function with parameters 
$(\Phi^*, M_V^*, \alpha) = (1.8\times 10^{-3}, -23.4, -1.85)$, while
the TVD simulation suggests a flatter faint-end slope of $\alpha\sim -1.2$. 
A comparison with $z=3$ shows that the rest-frame $V$-band 
luminosity function has brightened by about 0.5 magnitude from 
$z=3$ to $z=2$ without a significant change in the shape.
Our results do {\it not} imply that hierarchical galaxy formation 
fails to account for the massive galaxies at $z\gtsim 1$.
\end{abstract}



\section{Introduction}
\label{section:intro}

A number of recent observational studies have revealed a new population
of red, massive galaxies at redshift $z\sim 2$ \citep[e.g.][]{Chen03, Daddi04,
Franx, Glazebrook04}, utilizing near-infrared (IR) wavelengths which are
relatively less affected by dust extinction.  At the same time, a number of
studies focused on the assembly of stellar mass density at high redshift by
comparing observational data and semi-analytic models of galaxy formation
\citep[e.g.][]{Poli, Fontana03, Dick03a}. These works argued that the
hierarchical structure formation theory may have difficulty in accounting for
sufficient early star formation.  These concerns grew with the
mounting evidence for high redshift galaxy formation including the
discovery of Extremely Red Objects (EROs) at $z\ge 1$
\citep[e.g.][]{Elston88, McCarthy92, Hu94, Cimatti03, Smail02},
sub-millimeter galaxies at $z\ge 2$ \citep[e.g.][]{Smail97, Chapman03},
Lyman break galaxies (LBGs) at $z\sim 3$ \citep[e.g.][]{Steidel99}, and
galaxies at $z\gtsim 4$, detected either by their Lyman-$\alpha$ emission
\citep[e.g.][]{Hu99, Rhoads, Taniguchi, Kodaira, Ouchi03a} or by their
optical to near-IR colors \citep[e.g.][]{Iwata, Ouchi03b, Dick03b, Chen04}.  
We now face the important question as to whether this evidence for
high-redshift galaxy formation is consistent with the concordance 
$\Lam$CDM model.

The redshift range around $z\simeq 2$ is a particularly interesting epoch
also for another reason.  The redshift interval $1.4<z<2.5$ has long been
known as the `redshift desert' \citep[][]{Abraham04, Steidel04} for 
galaxy surveys, because there was no
large volume limited sample of galaxies available in this regime until very
recently.  This is because strong emission lines from H {\sc ii} regions of
galaxies, such as [O {\sc ii}] $\lam$3727, [O {\sc iii}] $\lam\lam$ 4959,
5007, H$\alpha$ and H$\beta$, redshift out of optical wavelengths above
9300~\AA. These spectral features are necessary to easily identify
galaxy redshifts, allowing ground-based telescopes to benefit from the
low-night sky background and high atmospheric transmission in the
wavelengths range 4000 -- 9000~\AA.

However, recently \citet{Ade04} and \citet{Steidel04} have introduced new
techniques for exploring the `redshift desert', making it possible to
identify a large number of galaxies efficiently with the help of
a color selection criteria in the color-color plane of $U_n -G$ vs. ~$G-R$.
In this technique, galaxies at $z=2 - 2.5$ are located 
photometrically from the mild drop in the $U_n$ filter owing to the 
Ly-$\alpha$ forest opacity, and galaxies at $z=1.5-2$ are recognized from 
the lack of a break in their observed-frame optical spectra.
The large sample of galaxies identified by these authors at $z=2$ makes it
now possible to study galaxy formation and evolution for over 10~Gyrs of
cosmic time, from redshift $z=3$ to $z=0$, without a significant gap.  We
note that the epoch around $z=2$ is particularly important for
understanding galaxy evolution because at around this time, the number
density of quasi-stellar objects (QSOs) peaked \citep[e.g.][]{Schmidt68, 
Schmidt95, Fan01b} and the ultra-violet (UV) luminosity density began 
its decline by about an order of magnitude from $z\sim 2$ to $z=0$ 
\citep[e.g.][]{Lilly96, Connolly97, Sawicki97, Treyer98, Pas98, Cowie99}.

These recent observational studies of galaxies at $z=2$, both in the UV and
near-IR wavelengths, imply a range of novel tests for the hierarchical
structure formation theory. In this study, we analyze the properties of
massive galaxies at $z=2$ formed in state-of-the-art cosmological
hydrodynamic simulations of the $\Lam$CDM model, and we compare them with
observations.

In an earlier recent study \citep{Nachos1}, we argued that, based on two
different types of hydrodynamic simulations (Eulerian TVD and SPH) and the
theoretical model of \citet{Her03} (hereafter H\&S model), the predicted
cosmic star formation rate (SFR) density peaks at $z\geq 5$, with a {\it
  higher} stellar mass density at $z=3$ than suggested by current
observations \citep[e.g.][]{Brinch, Cole, Cohen, Dick03a, Fontana03,
  Rudnick03, Glazebrook04}, in contrast to some claims to the contrary
\citep[][]{Poli, Fontana03}.  We also compared our results with those from
the updated semi-analytic models of \citet{Som}, \citet{Granato00}, and
\citet{Menci02}, and found that our simulations and the H\&S model predicts
an earlier peak of the SFR density and a faster development of stellar mass
density compared to these semi-analytic models.

It is then interesting to examine what our simulations predict for the
properties of massive galaxies at $z\sim 2$. In this paper, we analyze for
this purpose the same set of hydrodynamic simulations that was used in
\citet{Nachos1}, with a special focus on the most massive galaxies at this
epoch.  In Section~\ref{sec:simulation}, we briefly describe the
simulations that we use. In Section~\ref{sec:method}, we summarize our
method for computing spectra of simulated galaxies both in the rest-frame
and the observed-frame.  In Section~\ref{sec:colcol}, we show the
color-color diagrams and color-magnitude diagrams of simulated galaxies.
In Section~\ref{sec:mstar} we discuss the stellar masses and number density 
of galaxies at $z=2$. We then investigate the observed-frame $R$-band
luminosity function and the rest-frame $V$-band luminosity function in
Section~\ref{sec:lf}, followed by an analysis of the star formation
histories in Section~\ref{sec:sfhistory}.  Finally, we summarize and
discuss the implications of our work in Section~\ref{sec:discussion}.


\section{Simulations}
\label{sec:simulation}

We will discuss results from two different types of cosmological 
hydrodynamic simulations.  Both approaches include ``standard'' 
physics such as radiative cooling/heating, star formation,
and supernova (SN) feedback, although the details of the models and
the parameter choices vary considerably.

One set of simulations was performed using an Eulerian approach, which
employed a particle-mesh method for the gravity and the total variation
diminishing (TVD) method \citep{Ryu93} for the hydrodynamics, both with a
fixed mesh. The treatment of the radiative cooling and heating is described
in \citet{Cen92} in detail. The structure of the code is similar to that of
\citet{CO92, CO93}, but the code has significantly improved over the years
with additional input physics. It has been used for a variety of studies,
including the evolution of the intergalactic medium \citep{CO94, CO99a,
  CO99b}, damped Lyman-$\alpha$ absorbers \citep{Cen03}, and galaxy
formation \citep*[e.g.][]{CO00, Nag01a, Nag01b, Nag02}.

Our other simulations were done using the Lagrangian smoothed particle
hydrodynamics (SPH) technique.  We use an updated version of the code
{\small{GADGET}} \citep{Gadget}, which uses an `entropy conserving'
formulation \citep{SH02} of SPH to mitigate problems with energy/entropy
conservation \citep[e.g.][]{Her93} and overcooling.  The code also employed
a subresolution multiphase model of the interstellar medium to describe
self-regulated star formation which includes a phenomenological model for
feedback by galactic winds \citep{SH03a}, and the impact of a uniform
ionizing radiation field \citep{KWH96, Dave99}.  This approach has been used to
study the evolution of the cosmic SFR \citep{SH03b}, damped Lyman-$\alpha$
absorbers \citep*{NSH04a, NSH04b}, Lyman-break galaxies \citep*{NSHM}, 
disk formation \citep{Robertson04}, emission from the intergalactic medium
\citep{Fetal03,Fetal04a, Fetal04b, Fetal04c, Fetal04d, Zetal04},
and the detectability of high redshift galaxies \citep{Bart04, Fetal04e}.

In both codes, at each time-step, some fraction of gas is converted into
star particles in the regions that satisfy a set of star formation criteria
(e.g. significantly overdense, Jeans unstable, fast cooling, converging gas
flow). Upon their creation, the star particles are tagged with physical
parameters such as mass, formation time, and metallicity. After forming,
they interact with dark matter and gas only gravitationally as
collisionless particles. 

More specifically, in the TVD simulation, the star formation rate is 
formulated as
\begin{equation}
\frac{\dd\rho_*}{\dd t} = c_* \frac{\rho_{\rm gas}}{t_*},
\label{eq:TVD}
\end{equation}
where $c_*$ is the star formation efficiency and $t_*$ is the star formation
time-scale.  In the TVD N864L22 run, $c_*=0.075$ and $t_*=\max(t_{\rm dyn},
10^7~\yr)$ were used, where $t_{\rm dyn}=\sqrt{3\pi/(32G\rho_{\rm tot})}$
is the local dynamical time-scale owing to gravity.  At each time-step, a part
of the gas in a cell is converted to a star particle according to
Equation~(\ref{eq:TVD}), provided the gas is (a) moderately overdense 
($\delta_{\rm tot}>5.5$), (b) Jeans unstable ($m_{\rm gas}>m_J$),
(c) cooling fast ($t_{\rm cool}<t_{\rm dyn}$), and (d) the flow is
converging into the cell ($\nabla\cdot {\bf v} < 0$).

On the other hand, the SPH simulations analyzed in this study parameterized
the star formation rate as
\begin{equation}
\frac{\dd\rho_*}{\dd t} = (1-\beta) \frac{\rho_c}{t_*},
\label{eq:SPH}
\end{equation}
where $\beta$ is the mass fraction of short-lived stars that 
instantly die as supernovae (taken to be $\beta=0.1$). 
The star formation time-scale $t_*$ is again taken to be proportional to 
the local dynamical time of the gas: 
$t_*(\rho) = t_0^* (\rho_{\rm th} / \rho)^{1/2}$, where the value of
$t_0^*=2.1 ~\Gyr$ is chosen to match the \citet{Kennicutt98} law.  In the
model of \citet{SH03a}, this parameter simultaneously determines a
threshold density $\rho_{\rm th}$, above which a multiphase structure of
the gas, and hence star formation, is allowed to develop.  This physical
density is $8.6\times 10^6 h^2\Msun\kpc^{-3}$ for the simulations in this
study, corresponding to a comoving baryonic overdensity of $7.7\times 10^5$
at $z=0$.
\citep[See][for a description of how $\rho_{\rm th}$ is
determined self-consistently within the model.]{SH03a}  Note that the density
$\rho_c$ that determines the star formation rate in equation~(\ref{eq:SPH})
is actually the average density of cold clouds determined by a simplified
equilibrium model of a multiphase interstellar medium, rather than the
total gas density. However, in star-forming regions, the density is so high
that the mean density in cold gas always dominates, being  close to
the total gas density. 

We will further discuss the differences between our star formation recipes 
in the two codes in some more detail in Section~\ref{sec:discussion}.  
The cosmological parameters adopted in the simulations are intended to 
be consistent with recent observational determinations 
\citep[e.g.][]{Spergel03}, as summarized in Table~\ref{table:simulation} 
where we list the most important numerical parameters of our primary runs.
While there are many similarities in the physical treatment between
the two approaches (mesh TVD and SPH), we see that the TVD simulation
has somewhat higher mass resolution, and the SPH somewhat higher 
spatial resolution. In that sense, the two approaches are complementary
and results found in common are expected to be robust.


\section{Analysis Method}
\label{sec:method}

We begin our analysis by identifying galaxies in the simulations as groups
of star particles. Since the two simulations use inherently different
methods for solving hydrodynamics, we employ slightly different methods for
locating galaxies in the two types of simulations, each developed to
suit the particular simulation method well. Because the stellar groups of
individual galaxies are typically isolated, there is little ambiguity
in their identification, such that systematic differences owing to group
finding methods are largely negligible.

For the TVD simulation, we use a version of the algorithm HOP \citep{HOP}
to identify groups of star particles.  This grouping method consists of two
discrete steps: in the first step, the code computes the overdensity at the
location of each particle, and in a second step, it then merges identified
density peaks based on a set of overdensity threshold parameters specified
by the user.  There are three threshold parameters: a peak overdensity
$\delta_{\rm peak}$ above which the density peak is identified as a
candidate galaxy, an outer overdensity $\delta_{\rm out}$ which defines the
outer contour of groups, and finally a saddle point overdensity
$\delta_{\rm sad}$ which determines whether two density peaks within the
outer density contour should be merged.  \citet{HOP} suggest using
$(\delta_{\rm out}, \delta_{\rm sad}, \delta_{\rm peak}) = (80, 200, 240)$
for identifying dark matter halos in N-body simulations.

We applied the algorithm to the star particles in our simulation using
various combinations of these parameters, and found that the results do not
depend very much on the detailed choice. This reflects the higher
concentration and isolation of star particle groups as compared to the dark
matter, making group finding typically straightforward.
However, the stellar peak
overdensities relative to the mean are so high that the saddle-overdensity
parameter in HOP is unimportant, and the small galaxies associated with a
large central galaxy at the center of a large dark matter halo are often
not well separated by this algorithm.  Therefore, we decided not to apply
the second step of the algorithm. Instead, we merge two density peaks
identified by the first step only if they are within two neighboring cells.
By visually inspecting the particle distribution of very massive dark matter 
halos, we found that this method succeeds in separating some of the small 
galaxies embedded in a large dark matter halo especially at the outskirts 
of the halo, and helps to somewhat mitigate the `overmerging problem' in
high density regions.  However, even with this revised method, the central
massive galaxy remains afterwards, and the addition of a small number of 
galaxies to the faint-end of the luminosity function does not change
the qualitative feature of our results. Clearly more work is needed 
to improve this situation on the grouping in the future, for example, 
using an improved grouping method such as {\small VOBOZ} \citep*{Neyrinck04}.

For the SPH simulations, we employ the same method as in \citet{NSHM},
where we identified isolated groups of star particles using a simplified
variant of the {\small SUBFIND} algorithm proposed by \citet{Springel01}.
In short, we first compute an adaptively smoothed baryonic density field
for all star and gas particles. This allows us to robustly identify centers
of individual galaxies as isolated density peaks.  We then find the full
extent of these galaxies by processing the gas and star particles in order
of declining density, adding particles one by one to the galaxies.  If all
of the 32 nearest neighbors of a particle have lower density, this particle
is considered to be a new galaxy `seed'. Otherwise, the particle is
attached to the galaxy that its nearest denser neighbor already belongs to.
If the two nearest denser neighbors belong to different galaxies, and one
of these galaxies has less than 32 particles, these galaxies are merged. If
the two nearest denser neighbors belong to different galaxies and both of
these galaxies have more than 32 particles, then the particle is attached
to the larger group of the two, leaving the other one intact.  Finally, we
restrict the set of gas particles processed in this way to those particles
which have at least a density of $0.01$ times the threshold density for
star formation, i.e.  $\rho_{\rm th}=8.6\times 10^6 h^2\Msun\kpc^{-3}$
\citep[see][for details on how the star formation threshold is
determined]{SH03a}.  Note that we are here not interested in the gaseous
components of galaxies -- we only include gas particles because they make
the method more robust owing to the improved sampling of the baryonic density
field.

We found that the above method robustly links star particles that belong to
the same isolated galaxy in SPH simulations. A simpler friends-of-friends
algorithm with a small linking length can achieve a similar result,
but the particular choice one needs to make for the linking length in this
method represents a problematic compromise, either leading to artificial
merging of galaxies if it is too large, or to loss of star particles that
went astray from the dense galactic core, if it is too small.  Note that,
unlike in the detection of dark matter substructures, no gravitational
unbinding algorithm is needed to define the groups of stars that make up
the galaxies formed in the simulations.  We consider only galaxies with at
least 32 particles (star and gas particles combined) in our subsequent
analysis.

For both simulations, each star particle is tagged with its mass, formation
time, and metallicity of the gas particle that it was spawned from.  Based
on these three tags, we compute the emission from each star particle, and
co-add the flux from all particles for a given galaxy to obtain the spectrum
of the simulated galaxy. We use the population synthesis model by
\citet{BClib03} which assumes a \citet{Chab03b} initial mass function (IMF)
within a mass range of $[0.1, 100]\,\Msun$, as recommended by
\citet{BClib03}.  Spectral properties obtained with this IMF are very
similar to those obtained using the \citet{Kroupa01} IMF, but the
\citet{Chab03b} IMF provides a better fit to counts of low-mass stars and
brown dwarfs in the Galactic disk \citep{Chab03a}. We use the high
resolution version of the spectral library of \citet{BClib03} which
contains 221 spectra describing the spectral evolution of a simple stellar
population from $t=0$ to $t=20$~Gyr over 6900 wavelength points in the
range of 91~\AA - 160~$\mu$m.

Once the intrinsic spectrum is computed, we apply the \citet{Calzetti00}
extinction law with three different values of $E(B-V)=0.0, 0.15, 0.3$ in
order to investigate the impact of extinction.  These values span the range
of extinction estimated from observations of LBGs at $z=3$ \citep{Ade00,
  Shapley01}. Using the spectra computed in this manner, we then derive
rest-frame colors and luminosity functions of the simulated galaxies.

To obtain the spectra in the observed frame, we redshift the spectra 
and apply absorption by the intergalactic medium (IGM) following 
the prescription of \citet{Madau95}.  Once the redshifted spectra 
in the observed frame are obtained, we convolve them with different
filter functions, including $U_n, G, R$ \citep{Steidel93} and 
standard Johnson bands, and compute the magnitudes in both AB and 
Vega systems.  Apparent $U_n$, $G$, $R$ magnitudes are computed in 
the AB system to compare our results with \citet{Ade04} and 
\citet{Steidel04}, while the rest-frame V-band magnitude is computed 
in the Vega system.


\section{Color-color \& color-magnitude diagrams}
\label{sec:colcol}

\begin{figure}
\epsscale{1.0}
\plotone{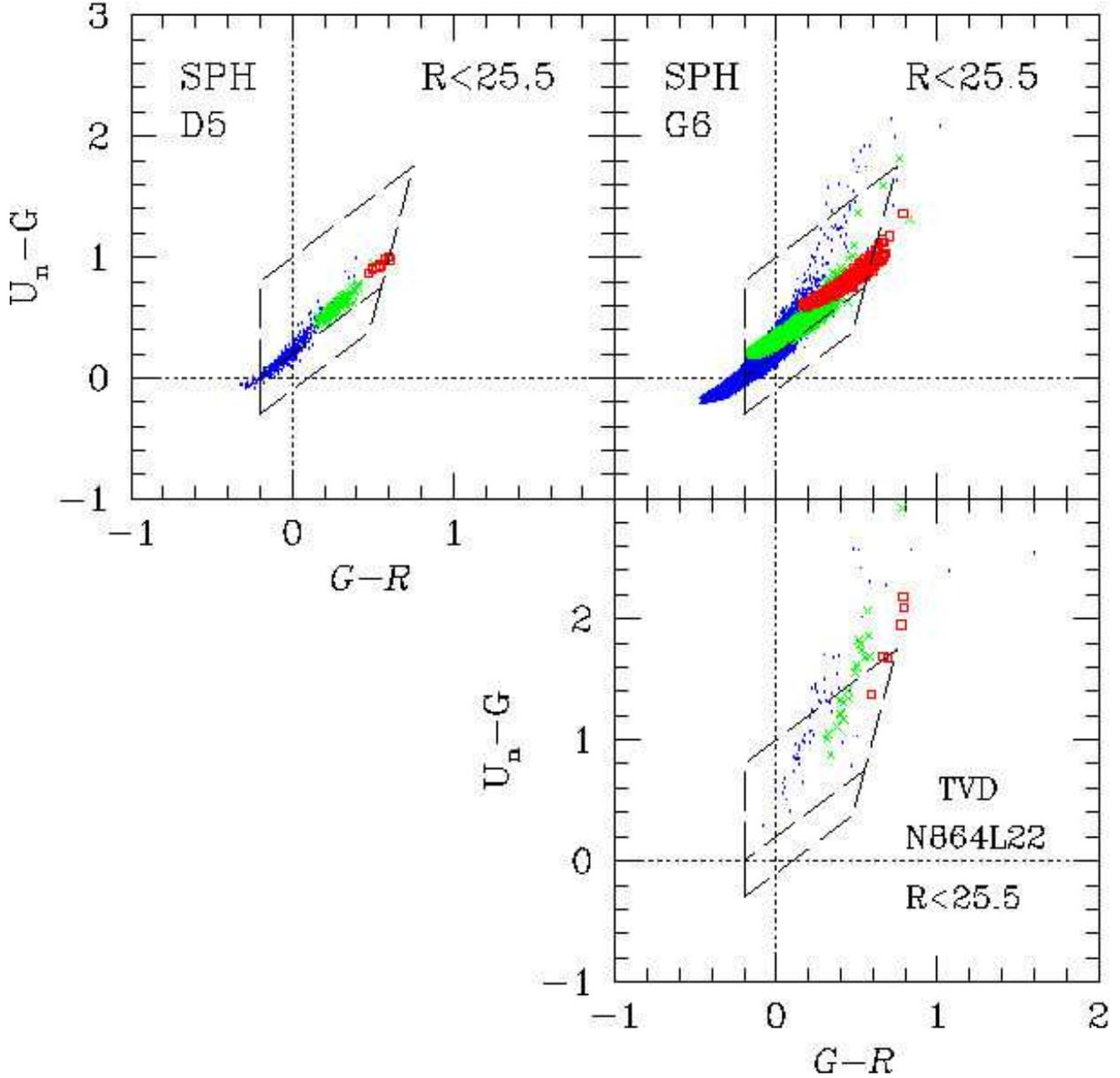}
\caption{Color-color diagrams of simulated galaxies at $z=2$ in the 
observed-frame $U_n - G$ vs. $G-R$ plane for both SPH (top 2 panels)
and TVD (bottom right panel) simulations. Only those galaxies that are 
brighter than $R=25.5$ are shown to match the magnitude limit 
of the \citet{Steidel04} sample. 
The three different symbols represent three different values of 
Calzetti extinction: $E(B-V)=0.0$
(blue dots), 0.15 (green crosses), and 0.3 (red open squares).
The number density of galaxies shown (i.e. $R<25.5$) is 
given in the bottom right corner of each panel for $E(B-V)=0.0$, 
0.15, 0.30 from top to bottom, respectively.
The color-selection criteria used by \citet{Ade04} and 
\citet{Steidel04} are shown by the long-dashed boxes. The upper 
(lower) box corresponds to their BX (BM) criteria for selecting out 
galaxies at $z=2.0-2.5$ ($z=1.5-2.0$).
}
\label{colcol.eps}
\end{figure}

In Figure~\ref{colcol.eps}, we show the color-color diagrams of simulated
galaxies at $z=2$ in the observed-frame $U_n - G$ vs. ~$G-R$ plane, for
both SPH (top 2 panels) and TVD (bottom right panel) simulations. Only
galaxies brighter than $R=25.5$ are shown in order to match the magnitude
limit of the sample of \citet{Steidel04}.  We plot different symbols for
three different values of Calzetti extinction: $E(B-V)=0.0$ (blue dots),
0.15 (green crosses), and 0.3 (red open squares).  As the level of
extinction by dust is increased from $E(B-V)=0.0$ to 0.3, the measured
points move towards the upper right corner of each panel. This behavior is
expected for a conventional star-forming galaxy spectrum, as demonstrated
in Figure 2 of \citet{Steidel03}.  The color-selection criteria used by
\citet{Ade04} and \citet{Steidel04} is shown by the long-dashed boxes. The
upper (lower) box corresponds to their BX (BM) criteria for selecting
galaxies at $z=2.0-2.5$ ($z=1.5-2.0$). The number density of all the
galaxies shown (i.e. $R<25.5$) is given in each panel for $E(B-V)=0.0$,
0.15, 0.30 from top to bottom, respectively.

It is encouraging to see that in all the panels most of the simulated
galaxies actually satisfy the observational color selection criteria for
the case of $E(B-V)=0.15$. This suggests that the simulated galaxies have
realistic UV colors compared to the observed ones.  Simulated galaxies with
no extinction tend to be too blue compared to the color-selection criteria.
In the SPH G6 run, the distribution is wider than in the SPH D5 and TVD
N864L22 runs, owing to the larger number of galaxies in a larger 
simulation box, and the distribution actually extends beyond the color 
selection boundaries. The spatial number density of galaxies with 
$R<25.5$ is about $2 \times 10^{-2}~h^3~\mpc^{-3}$ for the SPH G6 run.  
We will discuss the number density of galaxies with various cuts in 
Figure~\ref{nden.eps} in more detail.

We repeated the same calculation with the Sloan Digital Sky Survey's $u, g,
r, i, z$ filter set, and found that the distribution of points hardly
changes, even when we exchange $U_n-G$ with $u-g$, and $G-R$ with $g-r$.
This means that searches for galaxies in the redshift range $1.4<z<2.5$ can
also substitute the Sloan filters $u, g, r$ when applying a color-selection
following the `BX' and `BM' criteria adopted by \citet{Ade04} and
\citet{Steidel04}.

\begin{figure}
\epsscale{1.0}
\plotone{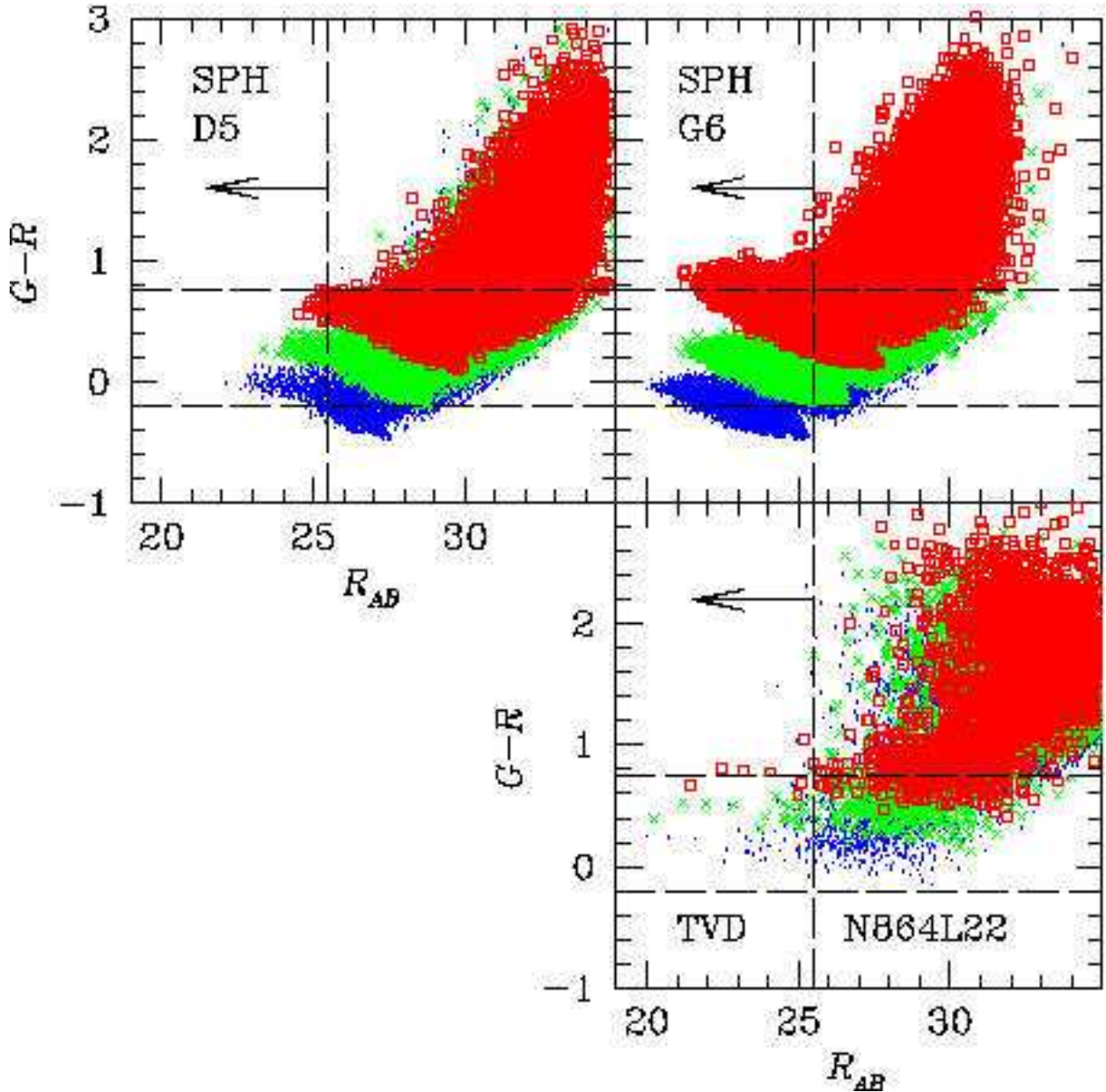}
\caption{Color-magnitude diagrams of simulated galaxies at $z=2$
on the $R$ vs. $G-R$ plane for both SPH (top 2 panels) and 
TVD (bottom right panel) simulations. All simulated galaxies are 
shown. Blue dots, green crosses, and red open squares correspond 
to $E(B-V)=0.0$, 0.15, 0.30. 
The magnitude limit of $R=25.5$ and the color range of 
$-0.2<G-R<0.75$ used by \citet{Steidel04} and \citet{Ade04} 
are shown by the long-dashed lines.
}
\label{colmag.eps}
\end{figure}

In Figure~\ref{colmag.eps}, we show the simulated galaxies at $z=2$ in the
plane of apparent $R$ magnitude and $G-R$ color. Again, we use three
different symbols for three different values of extinction: $E(B-V)=0.0$
(blue dots), 0.15 (green crosses), and 0.3 (red open squares). The vertical
long-dashed line and the arrow indicate the magnitude limit of $R=25.5$
used by \citet{Steidel04}. The horizontal line at $G-R=0.75$ corresponds to
the upper limit of the color-selection box of
\citet{Steidel04} and \citet{Ade04}.

We see that most of the galaxies brighter than $R=25.5$ automatically
satisfy the criterion $G-R<0.75$, and almost no galaxies
with $R<25.5$ fall out of the region for $E(B-V)=0.15$.  There is a
significant population of dim ($R>27$) galaxies with $G-R>1$. We will see
below that these are low-mass galaxies with stellar masses $\Mstar \le
10^{10}\himsun$.  We will discuss the amount of stellar mass contained in
the galaxies that satisfy the color selection criteria in
Section~\ref{sec:mstar}.


\section{Galaxy stellar masses at $z=2$}
\label{sec:mstar}

\begin{figure}
\epsscale{1.0}
\plotone{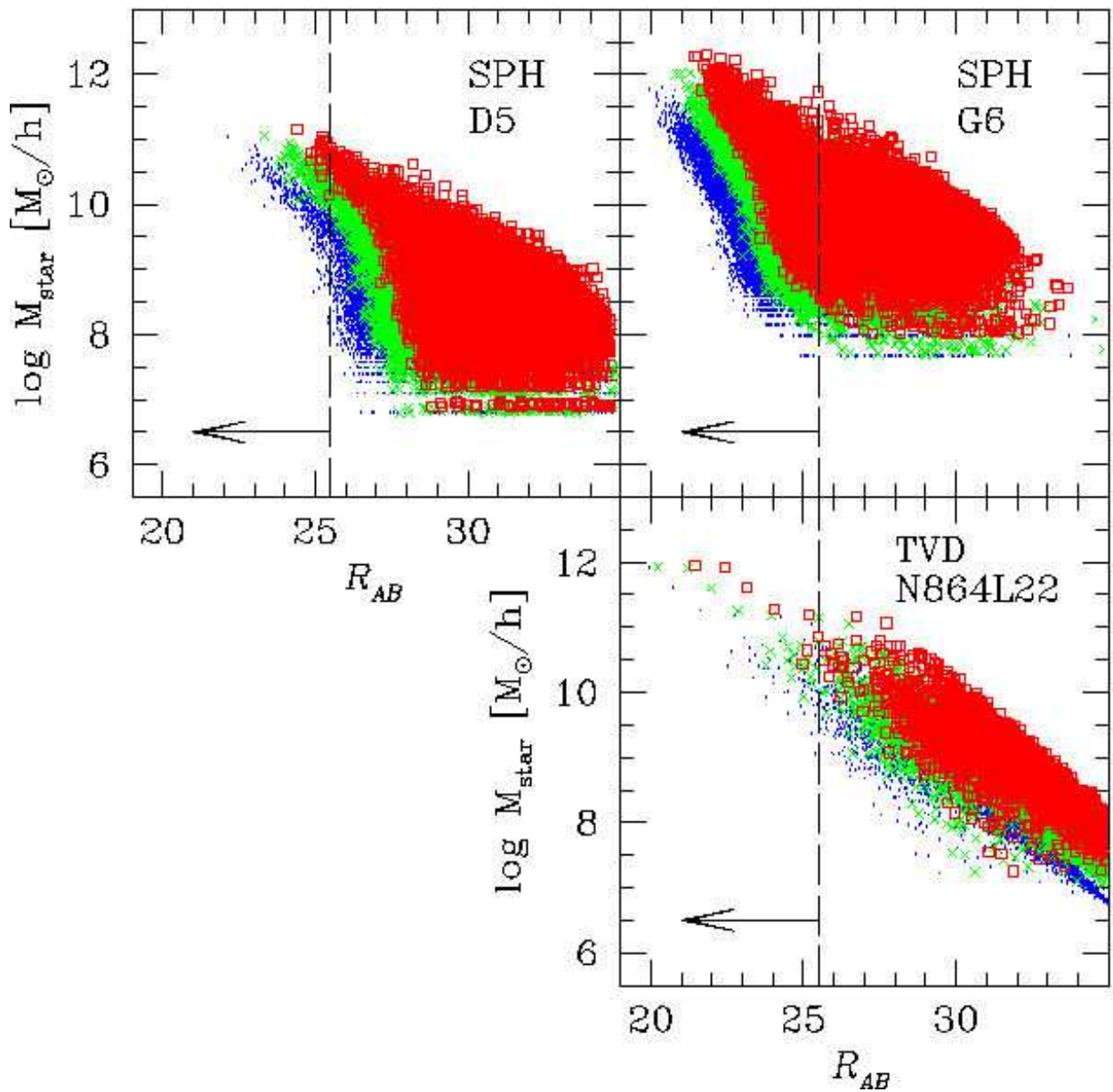}
\caption{Observed-frame $R$ magnitude vs. stellar masses of 
simulated galaxies at $z=2$ for both SPH (top 2 panels) and 
TVD (bottom right panel) simulations. Blue dots, green crosses, 
and red open squares correspond to $E(B-V)=0.0$, 0.15, 0.30. 
The magnitude limit of $R=25.5$ used by \citet{Steidel04} 
is shown by the vertical long-dashed line.
}
\label{Rmag_mstar.eps}
\end{figure}

\begin{figure}
\epsscale{1.0}
\plotone{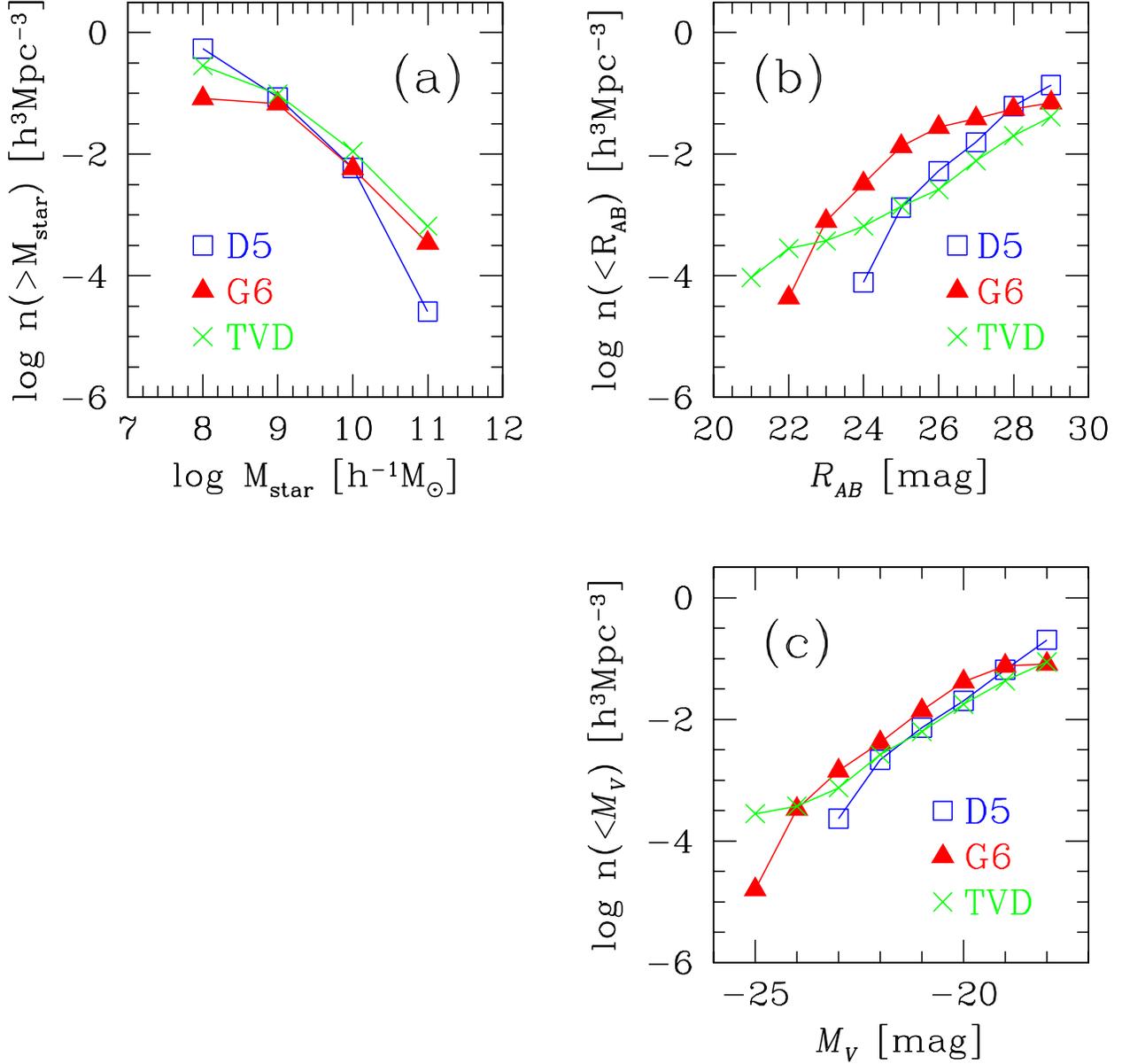}
\caption{Number density of all galaxies with stellar mass 
[{\it panel (a)}], observed-frame $R$-band magnitude 
[{\it panel (b)}], rest-frame $V$-band magnitude [{\it panel (c)}] 
above a certain value on the abscissa. The symbols correspond to 
the results from the SPH D5 run (blue open squares), the G6 run (red filled 
triangles), and the TVD run (green crosses).
}
\label{nden.eps}
\end{figure}

In Figure~\ref{Rmag_mstar.eps}, we show the apparent $R_{AB}$ magnitude
vs.~stellar mass of simulated galaxies at redshift $z=2$, for both SPH (top
2 panels) and TVD (bottom right panel) simulations.  The three different
symbols correspond to three different values of extinction: $E(B-V)=0.0$
(blue dots), 0.15 (green crosses), and 0.3 (red open squares). The vertical
long-dashed line and the arrow indicate the magnitude limit of $R=25.5$
used by \citet{Steidel04}.

From this figure, we see that there are many galaxies with stellar masses
larger than $10^{10}\himsun$ and $R<25.5$ at $z=2$ in our simulations.  In
the SPH G6 run, the masses of the most luminous galaxies are substantially
larger than those in the D5 run. But this is primarily a result of a finite
box size effect; the brightest galaxies have a very low space-density, and
so they are simply not found in simulations with too small a volume.  In the
SPH G6 run, the most massive galaxies at $z=2$ have masses $10^{11} <
\Mstar < 10^{12} \himsun$ with a number density of $3.5\times 10^{-4} h^3
\mpc^{-3}$. Such galaxies are hence not commonly found in a comoving
box-size of $\sim (30\mpc)^3$, in fact, only one such galaxy exists in the
D5 run.  Cosmic variance hence significantly affects the resulting number
density, and one ideally needs a simulation box larger than $\simeq
(100\himpc)^3$ to obtain a reliable estimate of the number density of such
massive galaxies.  We note however that a couple of such massive galaxies
are also found in the TVD N864L22 run, which is presumably in part a result
of overmerging owing to the relatively large cell size compared to the
gravitational softening length of the D5 run.  

In Figure~\ref{nden.eps}, we summarize the results on the cumulative number
densities of galaxies, either measured above a certain threshold value of
stellar mass [panel (a)], or below a threshold value of observed-frame
$R$-band magnitude [panel (b)], or rest-frame $V$-band magnitude [panel
(c)].  In panel (a), we see that the results of different runs agree
reasonably well in the threshold mass range of $10^9 < \Mstar
<10^{10}\himsun$, but differ somewhat at the lower and higher mass
thresholds.  At the low-mass end, the lower resolution of the G6 run
results in a lower number density than in the D5 run.  At the high-mass
end, the smaller box-size limits the number density in the D5 run compared
to the higher value in the G6 run.  The result of the TVD run falls in
between those of D5 and G6 at the low-mass end, but gives larger values
relative to the SPH runs at the high-mass end, probably owing to the
overmerging problem.  A similar level of agreement can be seen in panel
(c), where the rest-frame magnitude is used for the $x$-axis, but
in panel (b), comparatively large differences are present partly owing to
the larger stretch in the abscissa.

We find that the amount of stellar mass contained in the galaxies that
satisfy the color-selection criteria is in general far from being close to
the total stellar mass.  
Owing to the limited box-size and resolution effects, it is somewhat
difficult to obtain an accurate estimate of the mass fraction above
some magnitude limit.  However, we list the stellar mass fractions as a 
reference for future work. 
In the G6 run, which has comparatively low mass resolution, the stellar 
mass fraction contained in galaxies that pass the criteria 
(i.e. $R<25.5$ \& color-selection) account for only (60, 65, 42)\% of all 
the stars, for extinctions of $E(B-V)=(0.0, 0.15, 0.3)$, respectively.  
These fractions are relatively large compared to other runs, because,
as we will see in Section~\ref{sec:lf}, the luminosity function in the G6 
run starts to fall short of galaxies near $R\sim 25.5$ compared to the 
D5 run owing to its limited resolution, and the G6 run lacks 
low-mass faint galaxies with $R>25.5$.
On the other hand, in the D5 run, the stellar mass fraction contained 
in galaxies that pass the criteria (i.e. $R<25.5$ \& color-selection) 
account for only (32, 17, 2)\%, and (16, 13, 1)\% of the stars in the 
TVD N864L22 run, for extinctions of $E(B-V)=(0.0, 0.15, 0.3)$, respectively.
These fractions are smaller compared to those of the G6 run, because
the D5 and TVD N864L22 runs lack massive bright galaxies owing to 
their smaller box sizes.   
If we relax the criteria from `$R<25.5$ \& color-selection' to 
'only $R<25.5$', then the corresponding stellar mass fractions increase 
to (75, 65, 44)\% for G6, (32, 17, 3)\% for D5, and (46, 34, 29)\% 
for TVD N864L22. In summary, we expect that the true stellar mass fraction
that satisfies the criteria lies somewhere between the results of 
G6 and D5 (and TVD N864L22) runs, which should be roughly $\sim 50\%$. 

This implies that the current surveys with a magnitude limit of 
$R\sim 25.5$ at best account for half of the stellar mass in the Universe.  
\citet{Nachos1} reached a similar conclusion based on different theoretical 
arguments that compared the results of our hydrodynamic simulations and 
the theoretical model of
\citet{Her03} with near-IR observations of galaxies \citep[e.g.][]{Cole,
  Dick03a, Fontana03, Rudnick03, Glazebrook04}.  
\citet{Franx} and \citet{Daddi04} suggested that the unaccounted-for
contribution to the stellar mass might be hidden in a red population 
of galaxies. Our simulation suggests that the missed stellar masses 
are mostly contained in the fainter ($R>25.5$) and redder galaxies.
But note that we have not considered large values of extinction with
$E(B-V)>0.3$. 

\begin{figure}
\epsscale{1.0}
\plotone{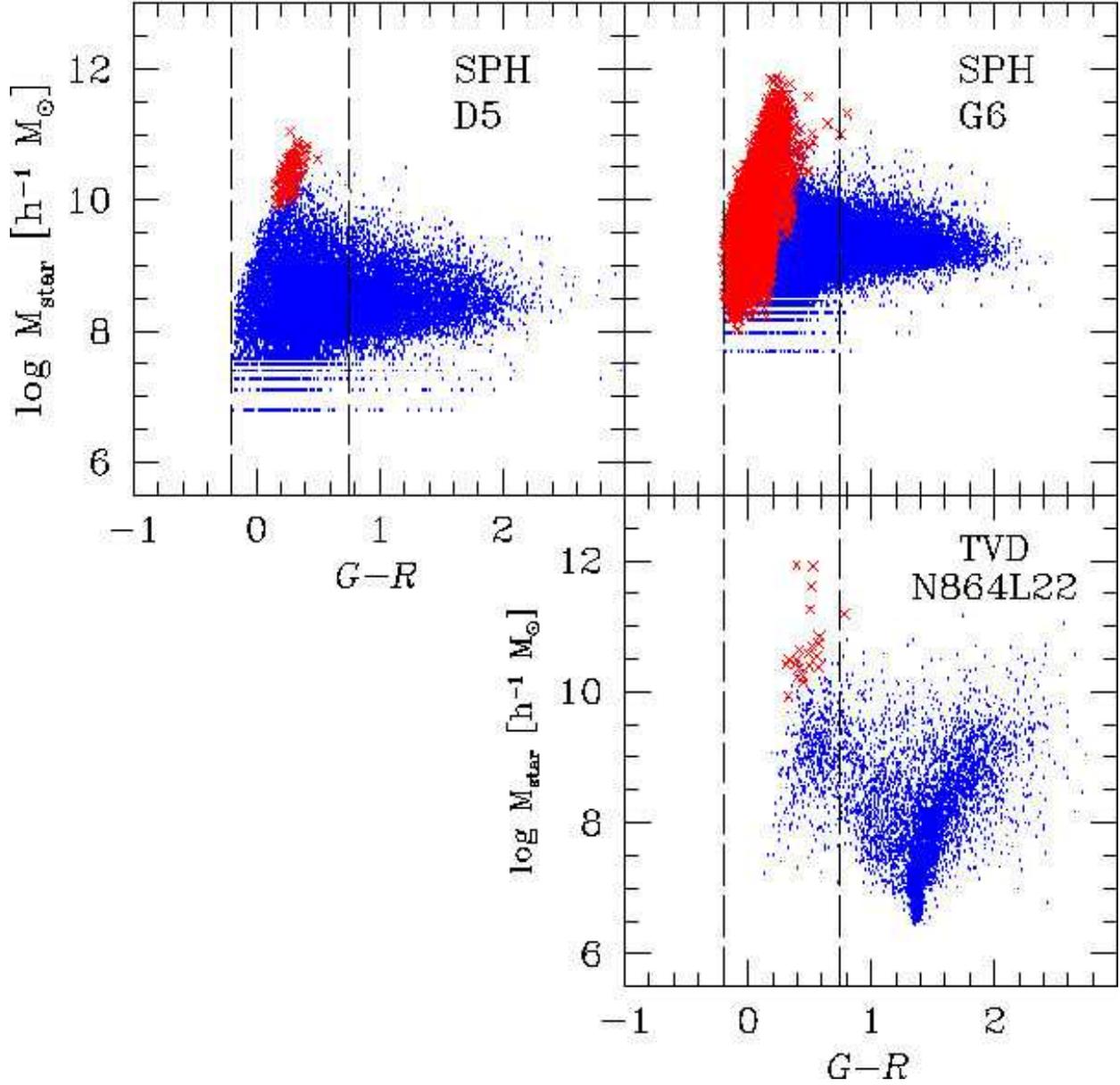}
\caption{Stellar masses vs. $G-R$ color of simulated galaxies at 
$z=2$ in the observed-frame for both SPH (top 2 panels) and 
TVD (bottom right panel) simulations. 
The red open crosses show the galaxies that are brighter than $R=25.5$, 
while the remainder of the points are shown in blue dots.
Here, only the case for $E(B-V)=0.15$ is shown.  
The most massive galaxies are not the bluest ones owing to the 
underlying old stellar population. The color range of $-0.2<G-R<0.75$, 
which is the $G-R$ color range of the selection criteria shown in 
Figure~\ref{colcol.eps}, is indicated by the vertical long-dashed lines. 
}
\label{mstar_col.eps}
\end{figure}

In Figure~\ref{mstar_col.eps}, we show the observed-frame $G-R$ color vs.
stellar mass of simulated galaxies at $z=2$ for both SPH (top 2 panels) and
TVD (bottom right panel) simulations.  Here, only the case of $E(B-V)=0.15$
is shown.  In the top two panels for the SPH runs, the red open crosses
show the galaxies that are brighter than $R=25.5$, and the blue dots
show the rest of the galaxies. The discrete stripes seen at low-mass end 
of the distribution for the SPH runs are due to the discreteness of the 
star particles in the simulation. For the TVD run, there is a strong 
concentration of points at $G-R \sim 1.4$, which we will discuss in detail
in Section~\ref{sec:sfhistory}.

An interesting point to note in this figure is that the bluest galaxies at
$z=2$ are not necessarily the most massive ones.  Rather, the bluest
galaxies with colors $-0.2<G-R<0.2$ are galaxies with somewhat lower mass
in the range $\Mstar < 10^{10}\himsun$. The most massive galaxies with
$\Mstar \gtsim 10^{10}\himsun$ have slightly redder colors of $0.2<G-R<0.6$
owing to their underlying old stellar population.

In the TVD simulation, there are a couple of galaxies with $G-R>2.0$
and $\Mstar > 10^{10} \himsun$, which are absent in the SPH simulations.
These galaxies are dominated by the old stellar population because 
they have completed their star formation before $z\sim 3$ as we will see 
in Section~\ref{sec:sfhistory}.


\section{Galaxy luminosity functions}
\label{sec:lf}

\begin{figure}
\epsscale{1.0}
\plotone{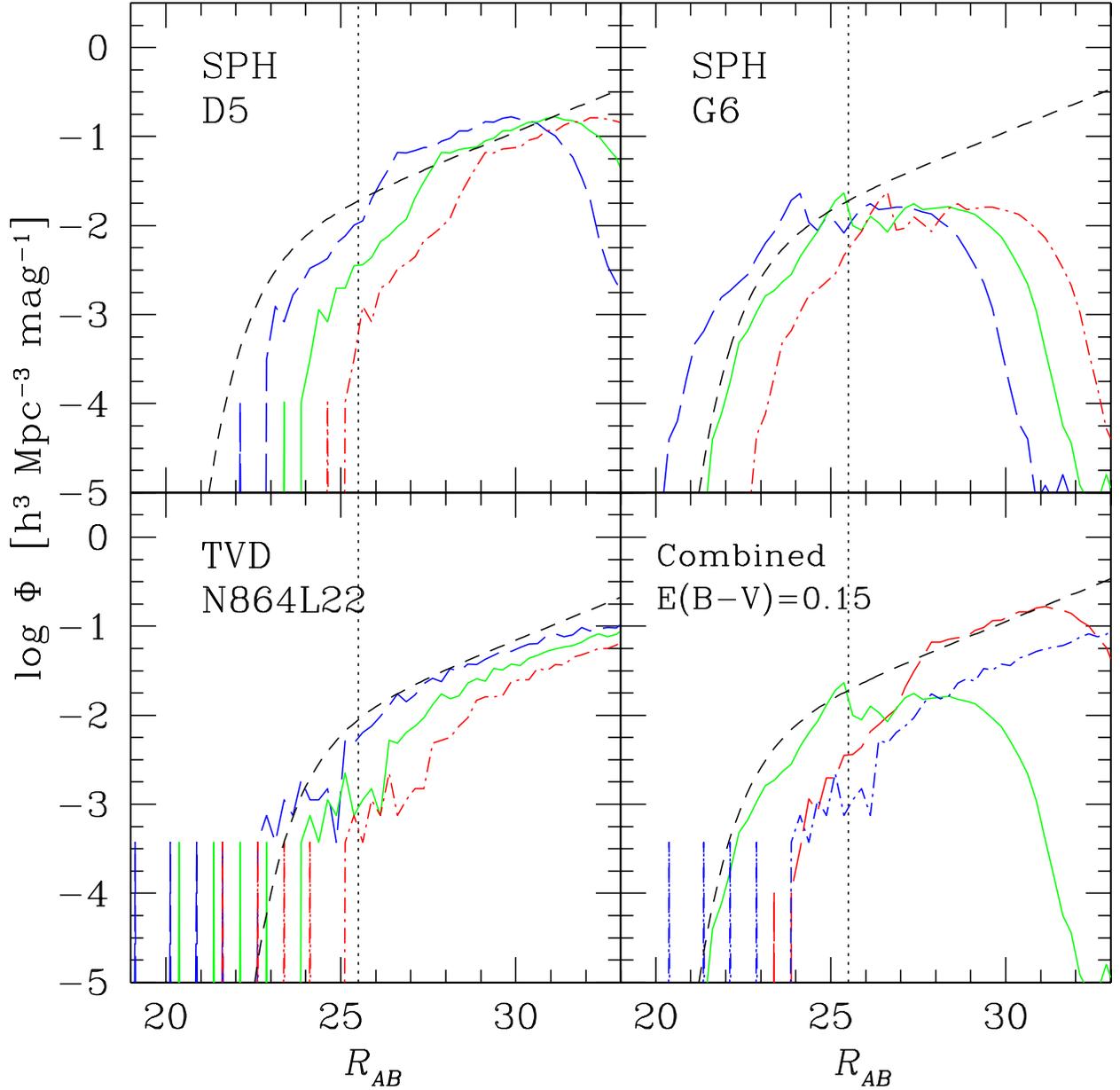}
\caption{$R$ band luminosity function in the observed-frame
for all simulated galaxies at $z=2$ for both SPH (top 2 panels) and 
TVD (bottom left panel) simulations. For these 3 panels, 
blue long-dashed, green solid, and red dot-short-dashed lines correspond
to $E(B-V)=0.0$, 0.15, 0.30, respectively. 
The vertical dotted line indicates the magnitude limit of 
$R=25.5$ used by \citet{Steidel04}. 
In the bottom right panel, a combined result is shown with the 
SPH D5 (green solid line), G6 (red long-dashed), and TVD (blue 
dot-dashed) results overplotted for the case of $E(B-V)=0.15$.
For comparison, the Schechter functions with following parameters 
are shown in short-dashed lines:
$(\Phi^*, M_R^*, \alpha) = (1\times 10^{-2}, 23.2, -1.4)$   
for SPH D5 \& G6 (top two panels), and combined results 
(bottom right panel), and $(1\times 10^{-2}, 24.5, -1.4)$ for 
the TVD run (bottom left panel), respectively. 
}
\label{lf_R.eps}
\end{figure}

Figure~\ref{lf_R.eps} shows the observed-frame $R$-band luminosity function
for all simulated galaxies at $z=2$ in both SPH (top 2 panels) and the TVD
(bottom left panel) simulations.  In these 3 panels, blue long-dashed,
green solid, and red dot-short-dashed lines correspond to $E(B-V)=0.0$,
0.15, 0.30, respectively.  The vertical dotted lines indicate the magnitude
limit of $R=25.5$ used by \citet{Steidel04}.  In the bottom right panel, a
combined result is shown with the SPH D5 (green solid line), G6 (red
long-dashed line), and TVD (blue dot-dashed line) results overplotted for
the case of $E(B-V)=0.15$.  For comparison, Schechter functions with the 
following parameters (simply chosen by eyeball fit) are shown as 
short-dashed lines: 
$(\Phi^*, M_R^*, \alpha) = (1\times 10^{-2}, 23.2,-1.4)$ 
for SPH D5 \& G6 runs (top two panels) and the combined results 
(bottom right panel), and 
$(1\times 10^{-2}, 24.5,-1.4)$ for TVD (bottom left), 
respectively. 

The bottom right panel of Figure~\ref{lf_R.eps} gives a relative comparison
between the different runs.  We see that the SPH G6 run contains a larger
number of luminous galaxies with $R<25$ compared to the other runs, an
effect that results from its larger box-size. On the other hand, the G6
run lacks fainter galaxies at $R>25$ compared to D5 run owing to
its limited
mass resolution. The SPH G6 run hence covers the bright-end of the
Schechter function with $M_R^*=23.2$, while the D5 run covers better the
faint-end of the luminosity function with a slope of $\alpha=-1.4$.  The
TVD N864L22 run agrees with the D5 results at the bright-end near $R=25$,
but the deviation from the D5 result starts to become somewhat large at
$R>28$. We will discuss a number of possible reasons for these differences
between the SPH and TVD runs in Section~\ref{sec:discussion}.

\begin{figure}
\epsscale{1.0}
\plotone{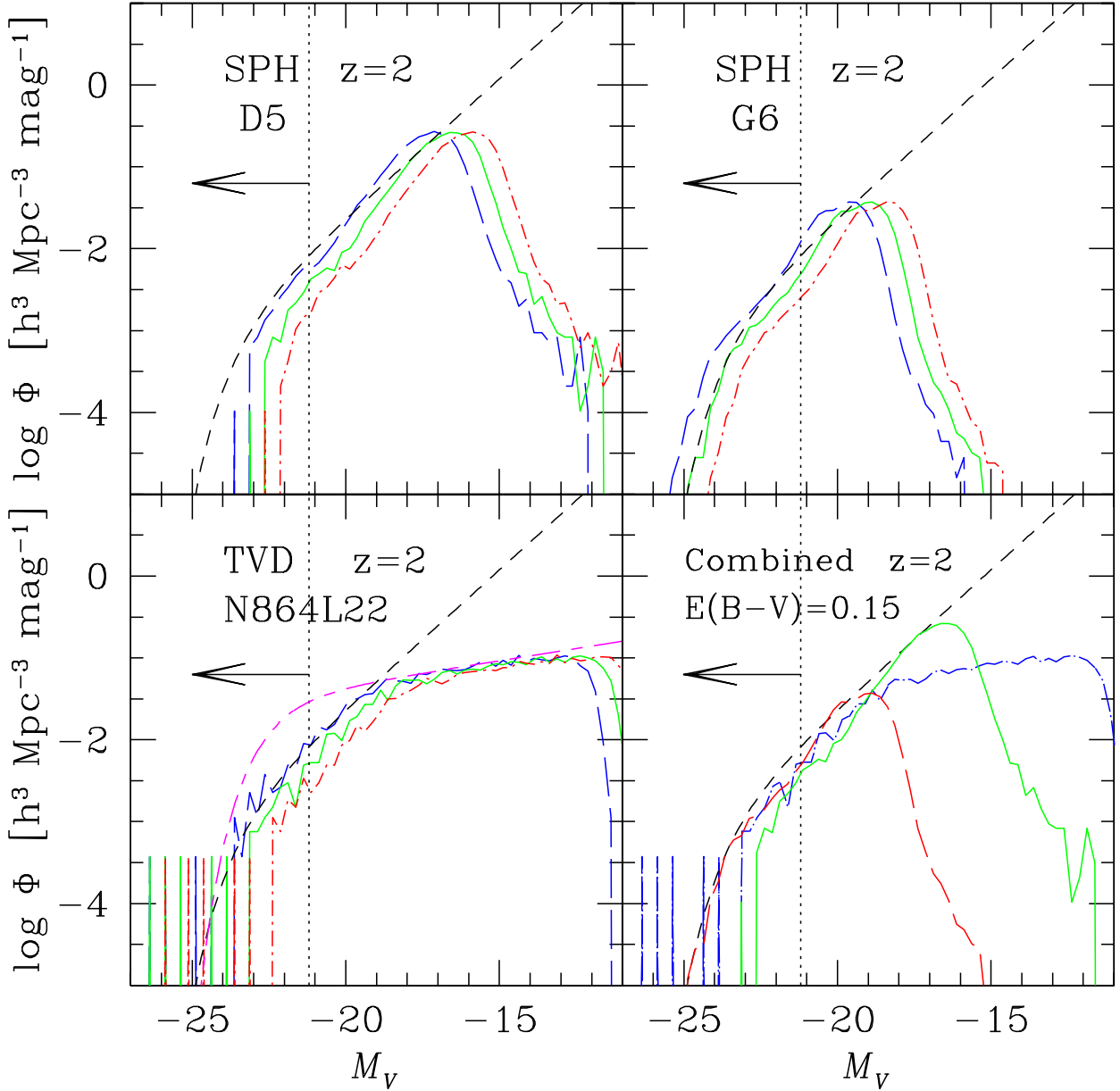}
\caption{Rest-frame $V$-band luminosity function of all simulated
galaxies at $z=2$ for both SPH (top 2 panels) and TVD (bottom left
panel) simulations. For these 3 panels, blue long-dashed, 
green solid, and red dot-short-dashed lines correspond to $E(B-V)=0.0$, 
0.15, 0.30, respectively. 
In the bottom right panel, a combined result is shown with the 
SPH D5 (green solid line), G6 (red long-dashed), and TVD (blue 
dot-dashed) results overplotted for the case of $E(B-V)=0.15$.
The magnitude limit of $M_V=-21.2$ (for $h=0.7$) for the survey of 
the LBGs at $z=3$ by \citet{Shapley01} is shown by the vertical dotted 
line. To guide the eye, a Schechter function with parameters 
($\Phi^*, M_V^*, \alpha) = (1.8\times 10^{-3}, -23.4, -1.8)$ is
shown in the short-dashed curve in all panels.
In the bottom left panel of TVD run, another Schechter function with
parameters 
($\Phi^*, M_V^*, \alpha) = (3.5\times 10^{-2}, -22.5, -1.15)$ is
also shown as a magenta long-dash-short-dashed line.
}
\label{lf_V.eps}
\end{figure}

In Figure~\ref{lf_V.eps}, we show the rest-frame $V$-band luminosity
function of all simulated galaxies at $z=2$ for both SPH (top 2 panels) and
TVD (bottom left panel) simulations. In these three panels, blue long-dashed,
green solid, and red dot-short-dashed lines correspond to extinctions of
$E(B-V)=0.0$, 0.15, 0.30, respectively.  In the bottom right panel, the
combined result is shown with the SPH D5 (green solid line), G6 (red
long-dashed), and TVD (blue dot-dashed) runs overplotted for the case of
$E(B-V)=0.15$.  For reference, the magnitude limit of $M_V=-21.2$ 
(for $h=0.7$)  for the survey of $z=3$ LBGs by \citet{Shapley01} is 
indicated by the vertical dotted line and the arrow. For comparison, 
we also include Schechter function fits with parameters 
($\Phi^*, M_V^*, \alpha) = (1.8\times 10^{-3}, -23.4, -1.8)$ as 
short-dashed curves in all panels.  These Schechter parameters are the 
same as the ones that \citet{Shapley01} fitted to their observational data, 
except that the characteristic magnitude $M_V^*$ is 0.5 magnitudes brighter.  
The same amount of brightening is seen in the G6 run from $z=3$ 
(see Fig. 5 of \citet{NSHM}\footnote{We note that there was an 
error in the calculation of the AB magnitudes in \citet{NSHM}. 
An additional factor of $5\log(1+z)=3.01$ has to be added to obtain 
correct AB magnitudes. The corrected version of the paper will be posted
at astro-ph/0311295. The rest-frame $V$-band results were not affected
by this error.}) to $z=2$. 
So the G6 run with $E(B-V)=0.15$ describes the bright-end 
of the luminosity function at $z=2$ and $z=3$ quite well.
In the bottom left panel for the TVD run, another Schechter function 
with a much shallower faint-end slope and parameters 
$(\Phi^*, M_V^*, \alpha) = (3.5\times 10^{-2}, -22.5, -1.15)$ is also 
shown with a magenta long-dash-short-dashed line for comparison. 
One can see that the TVD run contains more fainter galaxies with 
$M_V > -15$ than the SPH runs. This could be due to the higher baryonic 
mass resolution and the additional small-scale power in the TVD run 
compared with the SPH runs as we will discuss more in 
Section~\ref{sec:sfhistory}.

We can compute the expected characteristic magnitude $M_R^*$ of $z=2$ 
galaxies from the observational estimate of $M_R^*$ at $z=3$.
The difference in the luminosity distance between $z=3$ and $z=2$
corresponds to about one magnitude for our adopted cosmology, and we 
saw in the previous paragraph that the rest-frame $V$-band luminosity 
function brightened by $\sim 0.5$ magnitude from $z=3$ to $z=2$. 
Therefore, we expect that $M_R^*$ should be brighter by about 1.5 magnitudes 
at $z=2$ compared to $z=3$. \citet{Ade00} reported $M_R^*=24.54$ 
for the observed-frame $R$-band luminosity function of the $z=3$ galaxies. 
Therefore we obtain $M_R^*\sim 23.0$ as the expected value for the 
$z=2$ galaxies.
The result of the SPH G6 run shown in Fig.~\ref{lf_R.eps} is reasonably 
close to this expected value.  It is also clear that the box size of 
the D5 and TVD runs are too small to reliably sample the brightest galaxies 
at $z=2$. This is also seen in Fig.~\ref{lf_V.eps} as a lack of 
brightest galaxies at $M_V<-23$ in the D5 and TVD runs.
This is because the brightest galaxies form in rare high density peaks
which are at the knots of the large-scale filaments, and one needs a 
sufficiently large volume to sample a fair number of such bright 
galaxies reliably. Since the comoving correlation length of LBGs at 
$z=3$ are $\sim 4\himpc$ \citep{Ade03}, one needs a box size larger than 
at least $\Lbox \gtsim 40 \himpc$ in order to have a fair sample of 
LBGs at $z=3$.
The simulated volume of the G6 run is $(100\himpc)^3$ comoving, and one of 
the face of the simulation box corresponds to 2.5 ${\rm deg}^2$ at $z=3$. 
The coverage area of the LBG survey of \citet{Steidel03} at $z=3$ is about 
0.4 ${\rm deg}^2$, and the depth of their survey is $\Delta z \simeq 0.5$
which corresponds to $\sim 500 \himpc$ comoving for our adopted 
flat-$\Lam$ cosmology. 
Therefore the volumes of the G6 simulation and the Steidel et al. survey 
are comparable. 


\section{Star formation history of galaxies}
\label{sec:sfhistory}

\begin{figure}
\epsscale{1.2}
\plottwo{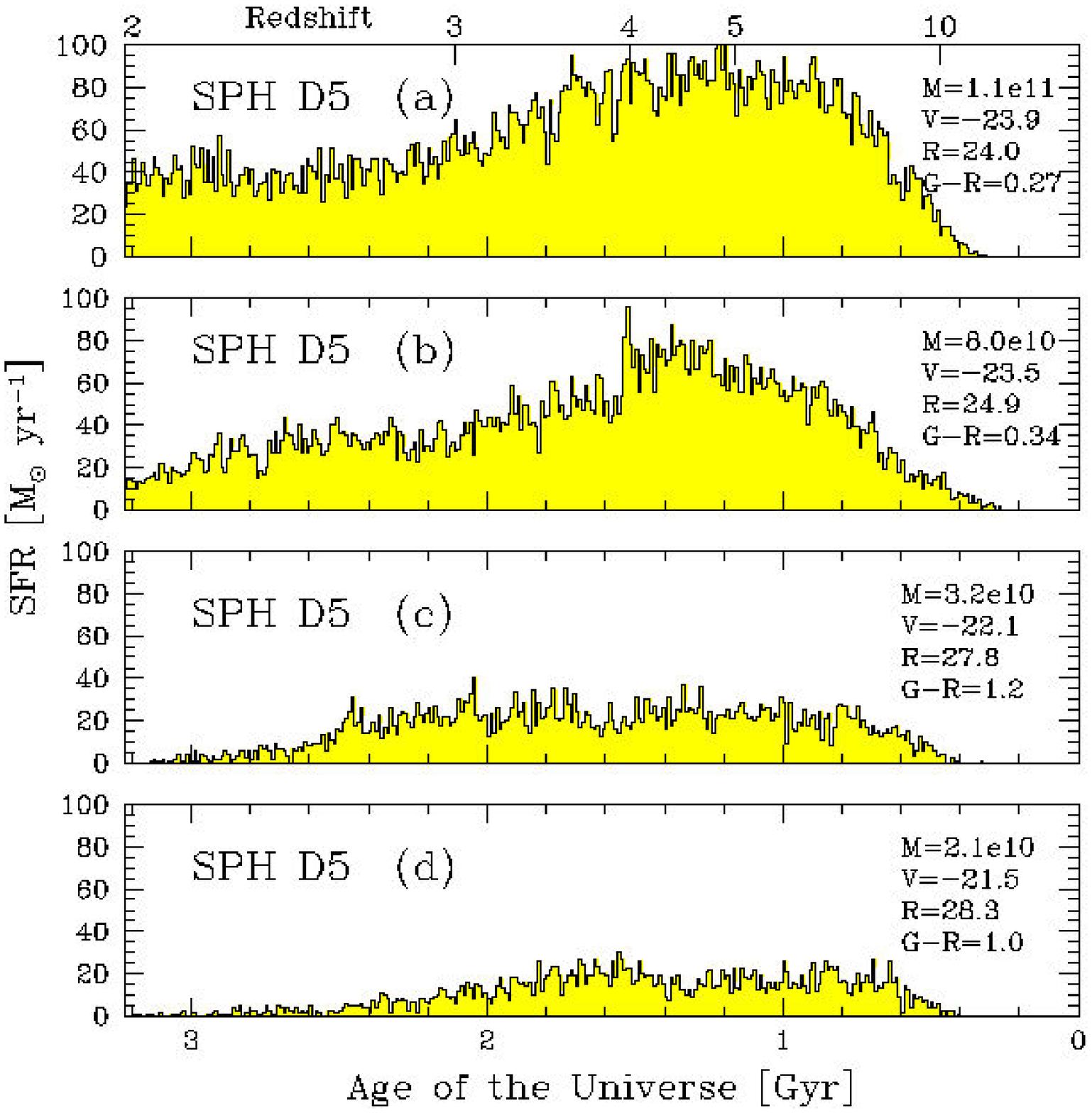}{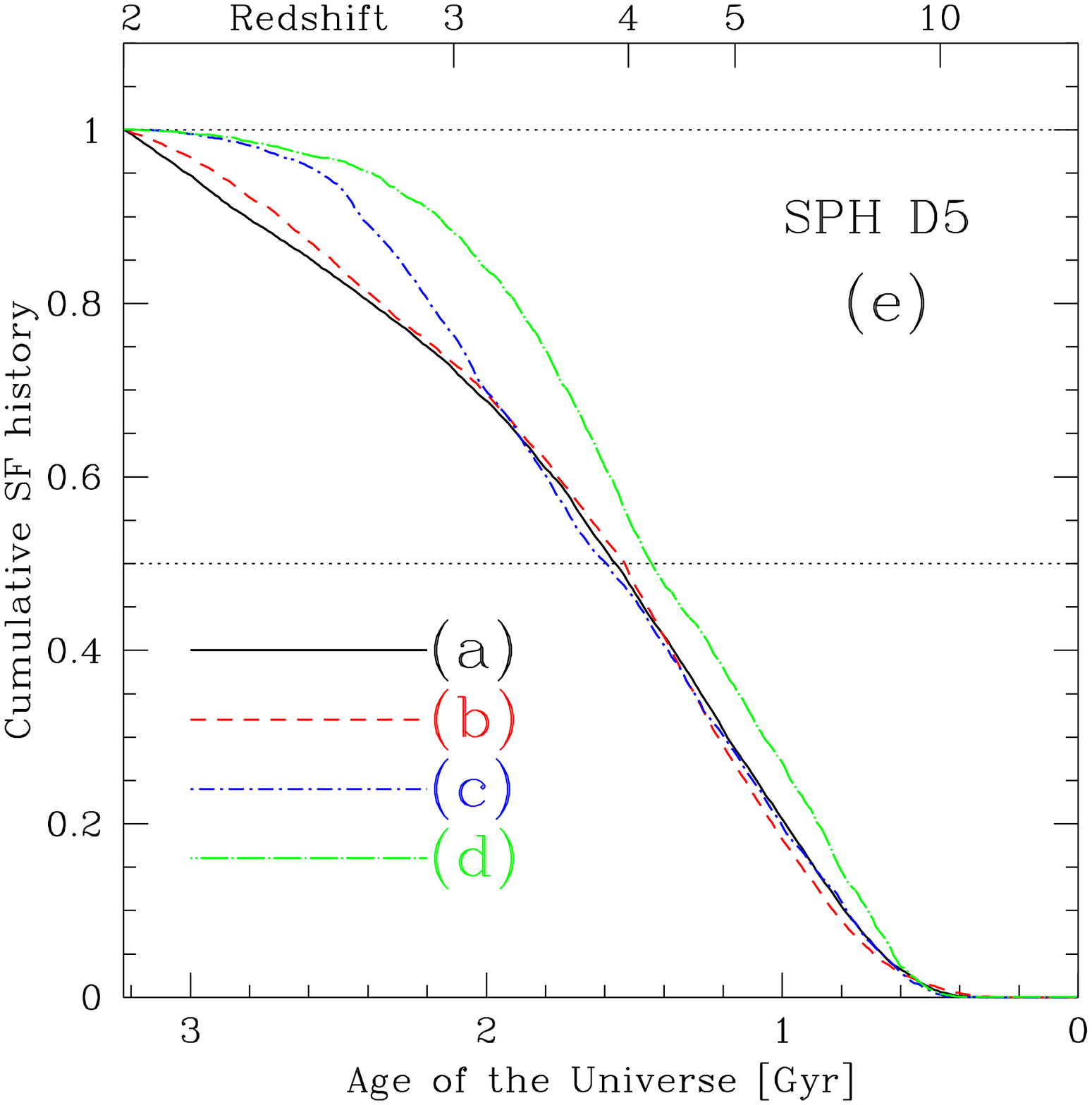}
\caption{
{\it Panels (a)  - (d)}: Star formation history as a function of 
cosmic time for selected galaxies in the SPH D5 run. 
In the right side of panels $(a)-(d)$, the following values for 
the case of $E(B-V)=0.15$ are listed: stellar mass
in units of $\himsun$, rest-frame V-band magnitude, observed-frame 
$R$-band magnitude, and $G-R$ color. 
Galaxies (a) \& (b) are the two most massive 
galaxies, and (c) \& (d) are the two reddest galaxies with 
$\Mstar > 1\times 10^{10}h^{-1}\Msun$. 
{\it Panel (e)}: Cumulative SF history of galaxies (a) - (d).
}
\label{sfhist_D5.eps}
\end{figure}

\begin{figure}
\epsscale{1.2}
\plottwo{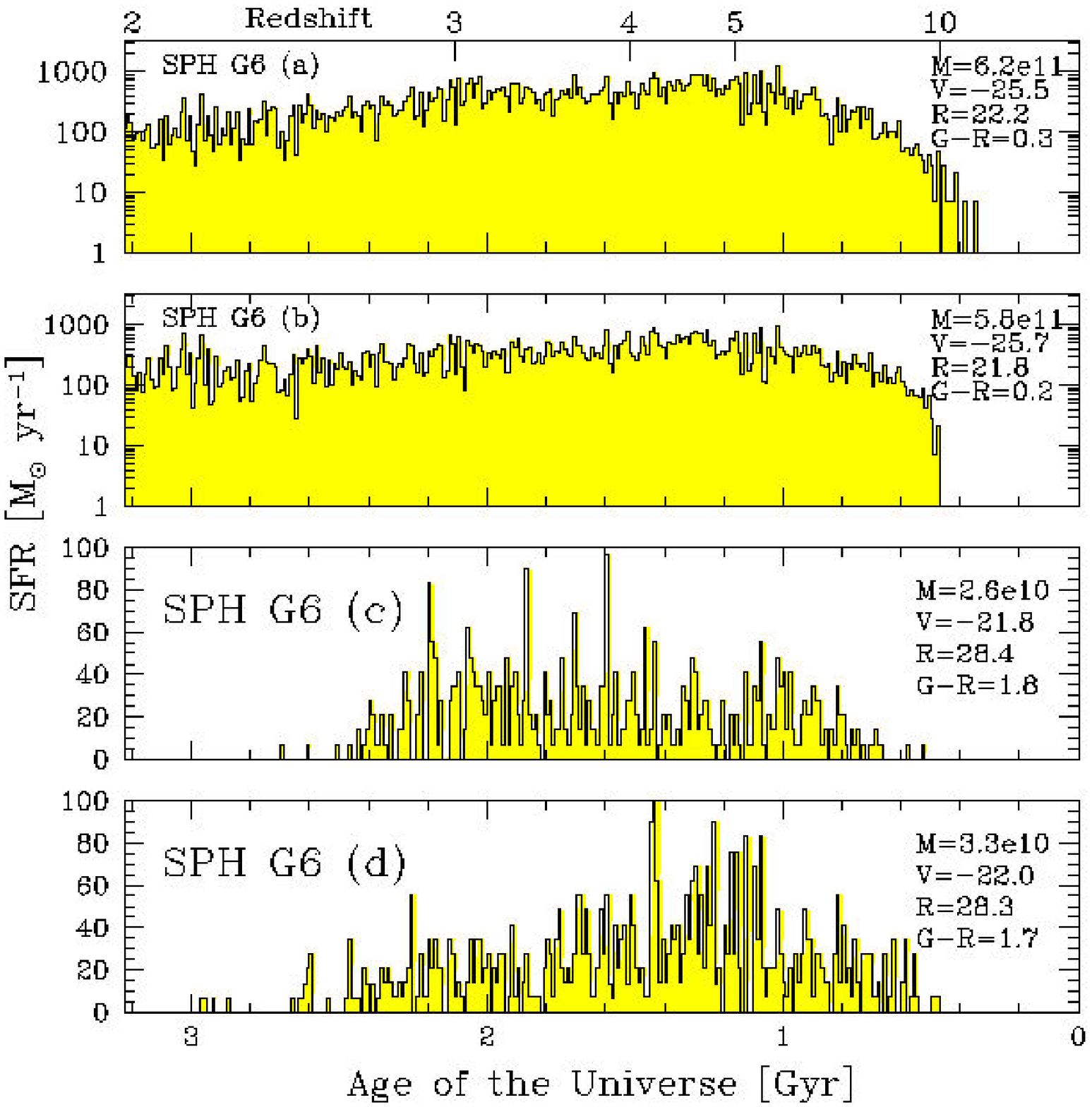}{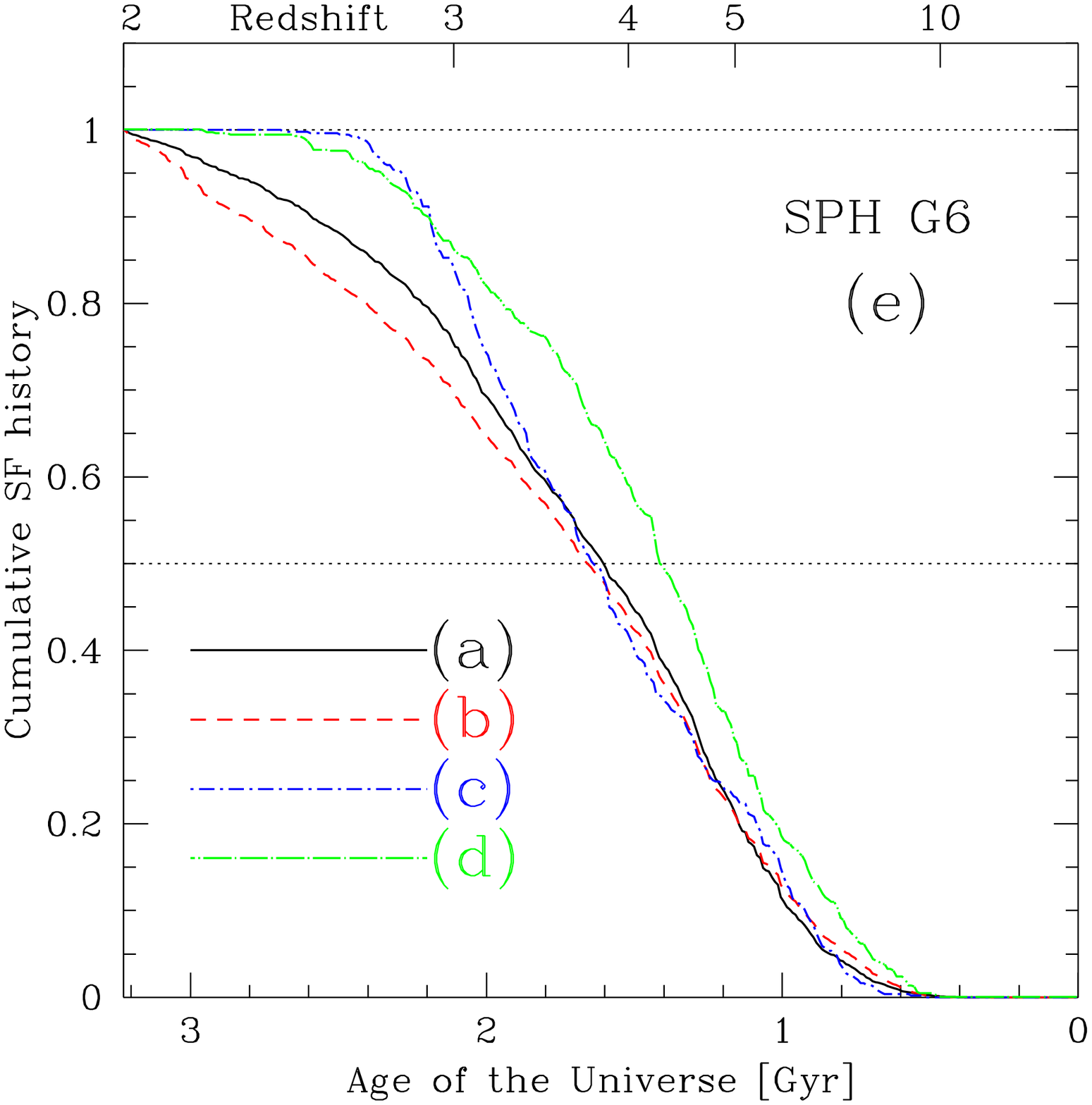}
\caption{
{\it Panels (a)-(d)}: Star formation history as a function of 
cosmic time for selected galaxies in
the SPH G6 run. In the right side of each panel,
the following values are listed: stellar mass in units of $\himsun$, 
rest-frame V-band magnitude, observed-frame $R$-band magnitude, 
and $G-R$ color. Galaxies (a) \& (b) are the two most massive 
galaxies, and (c) \& (d) are the two reddest galaxies with 
$\Mstar > 1\times 10^{10}h^{-1}\Msun$. Note that panels (a) \& (b) 
use a logarithmic scale for the ordinate, while panels (c) \& (d)
use a linear scale.
{\it Panel (e)}: Cumulative SF history of galaxies (a) - (d).
}
\label{sfhist_G6.eps}
\end{figure}

\begin{figure}
\epsscale{1.2}
\plottwo{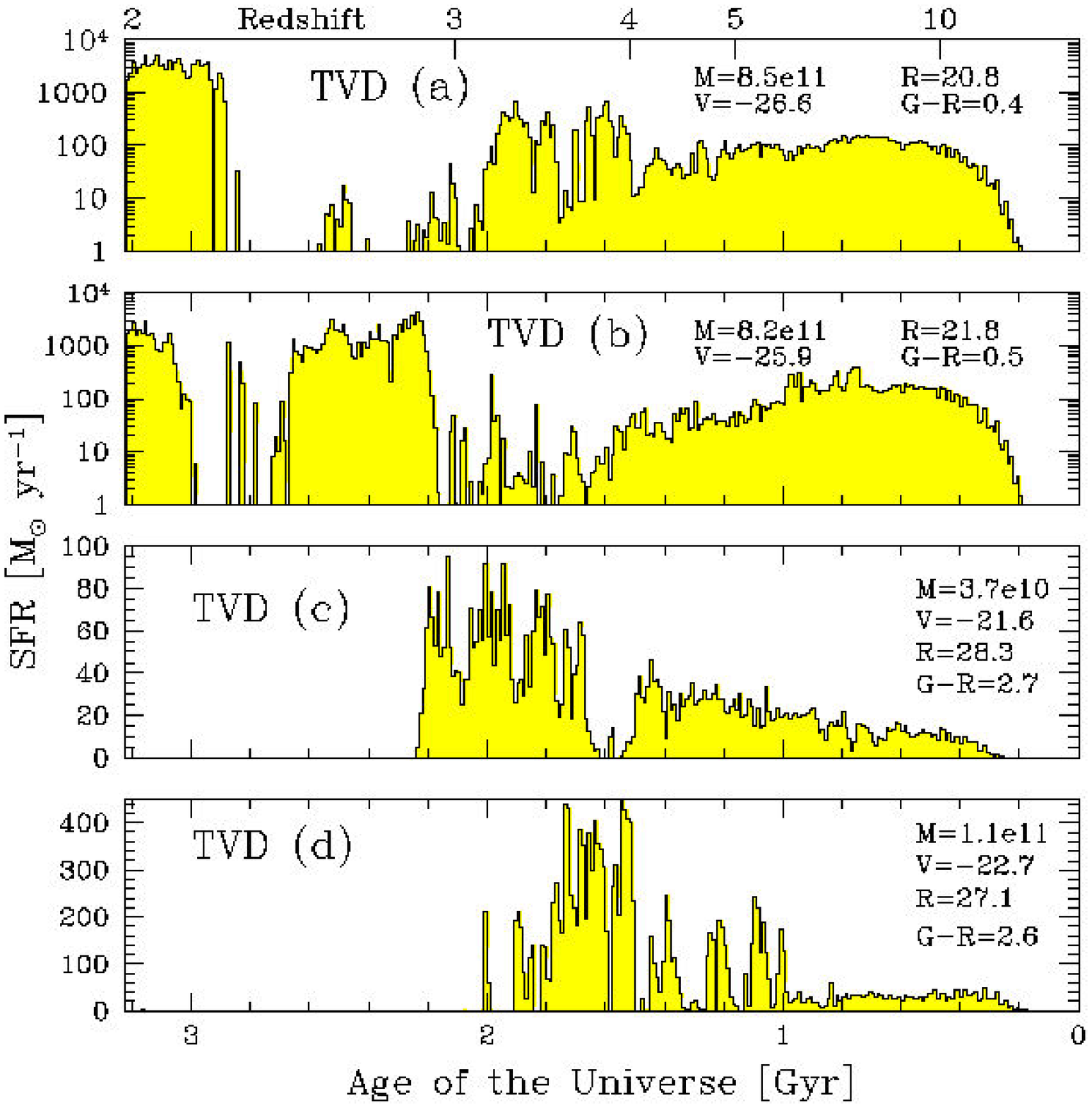}{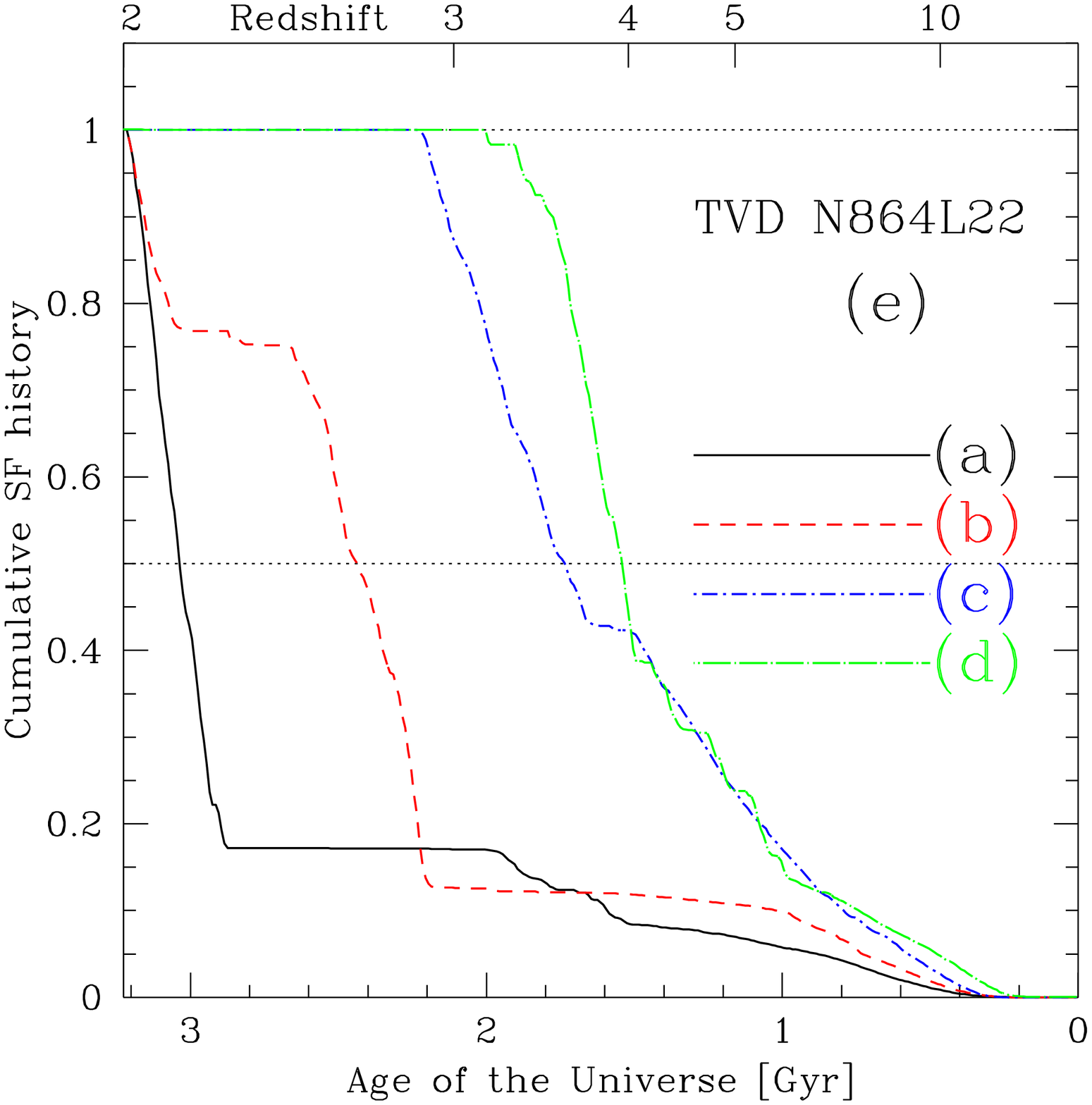}
\caption{
{\it Panels (a)-(d)}: Star formation history as a function of 
cosmic time for selected galaxies in
the TVD N864L22 run. In each panel, 
the following values are listed: stellar mass in units of $\himsun$, 
 rest-frame V-band magnitude, observed-frame $R$-band magnitude, 
and $G-R$ color. Galaxies (a) \& (b) are the two most massive 
galaxies, and (c) \& (d) are the two reddest galaxies with 
$\Mstar > 1\times 10^{10}h^{-1}\Msun$. Note that panels (a) \& (b) 
use a logarithmic scale for the ordinate, while panels (c) \& (d)
use a linear scale.
{\it Panel (e)}: Cumulative SF history of galaxies (a) - (d).
}
\label{sfhist_TVD.eps}
\end{figure}

In Figures~\ref{sfhist_D5.eps}, \ref{sfhist_G6.eps}, and
\ref{sfhist_TVD.eps}, we show the star formation (SF) histories of massive
galaxies at $z=2$ as a function of cosmic age, for the SPH D5,
G6, and TVD runs, respectively. In these figures, panels $(a)$ \& $(b)$
show results for the two most massive galaxies in each simulation box,
while panels $(c)$ \& $(d)$ are for the two reddest galaxies among those
with $\Mstar>10^{10}\himsun$.  In the right side of each panel, the
following quantities for the case of $E(B-V)=0.15$ are listed from 
top to bottom: 
stellar mass in units of $\himsun$, rest-frame V-band magnitude, 
observed-frame $R$-band magnitude, and $G-R$ color.  In panel $(e)$, 
the cumulative star formation history of the galaxies in panels 
$(a)-(d)$ is shown.

The results of the SPH D5 and G6 runs (Fig. \ref{sfhist_D5.eps} \&
\ref{sfhist_G6.eps}) suggest that the most massive galaxies have almost
continuously formed stars with a rate exceeding $30~\Msun~\yr^{-1}$ over a
few Gyrs from $z=10$ to $z=2$ (note that $30~\Msun~\yr^{-1} \times 3~\Gyr
\sim 10^{11}\Msun$).  Half of their stellar mass was already assembled by
$z=4$.  The reddest galaxies (panels $(c)$ \& $(d)$) have colors of $G-R
\gtsim 1.0$, and their star formation has become less active after $z=3$,
with the bulk of their stellar mass created before $z=3$ as can be seen 
in the panel $(e)$.

The TVD simulation (Fig.~\ref{sfhist_TVD.eps}) on the other hand suggests a
more sporadic star formation history than the SPH simulations, and the two
most massive galaxies in the simulation box experienced massive star
formation between $z=3$ and $z=2$ that exceeded $1000~\Msun~\yr^{-1}$ for
brief periods. The peak star formation rate of the most massive galaxy in 
the G6
run also reaches $1000~\Msun~\yr^{-1}$ occasionally.  In the real universe,
such violent starburst activity is found in high-redshift quasars and 
submm sources where the existence of dense, warm, and massive molecular 
clouds of $10^{10} - 10^{11} \Msun$ is inferred from the observed 
rotational emission line of CO \citep[e.g.][]{Omont03, Beelen04}.  
The number density of quasar host galaxies with such a violent star 
formation activity is not
well-constrained yet, but it probably is not very large because the space
density of quasars with $M_B<-26$ is smaller than $n_{\rm Q} \sim 10^{-6}
\mpc^{-3}$ \citep[e.g.][]{Pei95}. The space density of submm sources
is also similar: $n_{\rm submm}\sim 6.5\times 10^{-6} \mpc^{-3}$ at 
$z\sim 2$ \citep{Chapman03}.  These values are much smaller than the space
density of Lyman break galaxies $n_{\rm LBG} \sim 4 \times 10^{-3} h^3
\mpc^{-3}$ \citep{Ade04}. Finding 3 such massive starbursts at $z=2$ 
in our TVD simulation box of $(22 \himpc)^3$ results in a number density 
of $\sim 3\times 10^{-4} h^3 \mpc^{-4}$, which may be too high compared to
the observed value.  It is also possible that the two galaxies shown in
Figure~\ref{sfhist_TVD.eps} are affected by the overmerging problem which
tends to merge galaxies in very high density regions owing to a lack of
spatial resolution. If these two massive galaxies were broken up into
several less massive galaxies, then the $1000~\Msun~\yr^{-1}$ of star
formation would be distributed to a few hundreds $\Msun~\yr^{-1}$ for
each galaxy.  

The two reddest massive galaxies in the TVD run 
(Fig.~\ref{sfhist_TVD.eps}c \& \ref{sfhist_TVD.eps}d) have cumulative 
star formation histories which are similar to those of the SPH runs, 
with half of their stellar mass already assembled by $z\sim 4$. 
They have completed their star formation by $z\sim 3$ and have been 
quiet from $z=3$ to $z=2$. Because they are dominated by an old 
stellar population, they have very red colors of $G-R>2.0$ and show 
up in the upper right corner of the TVD (bottom right) panel in 
Fig.~\ref{colcol.eps}.

\begin{figure}
\epsscale{1.0}
\plotone{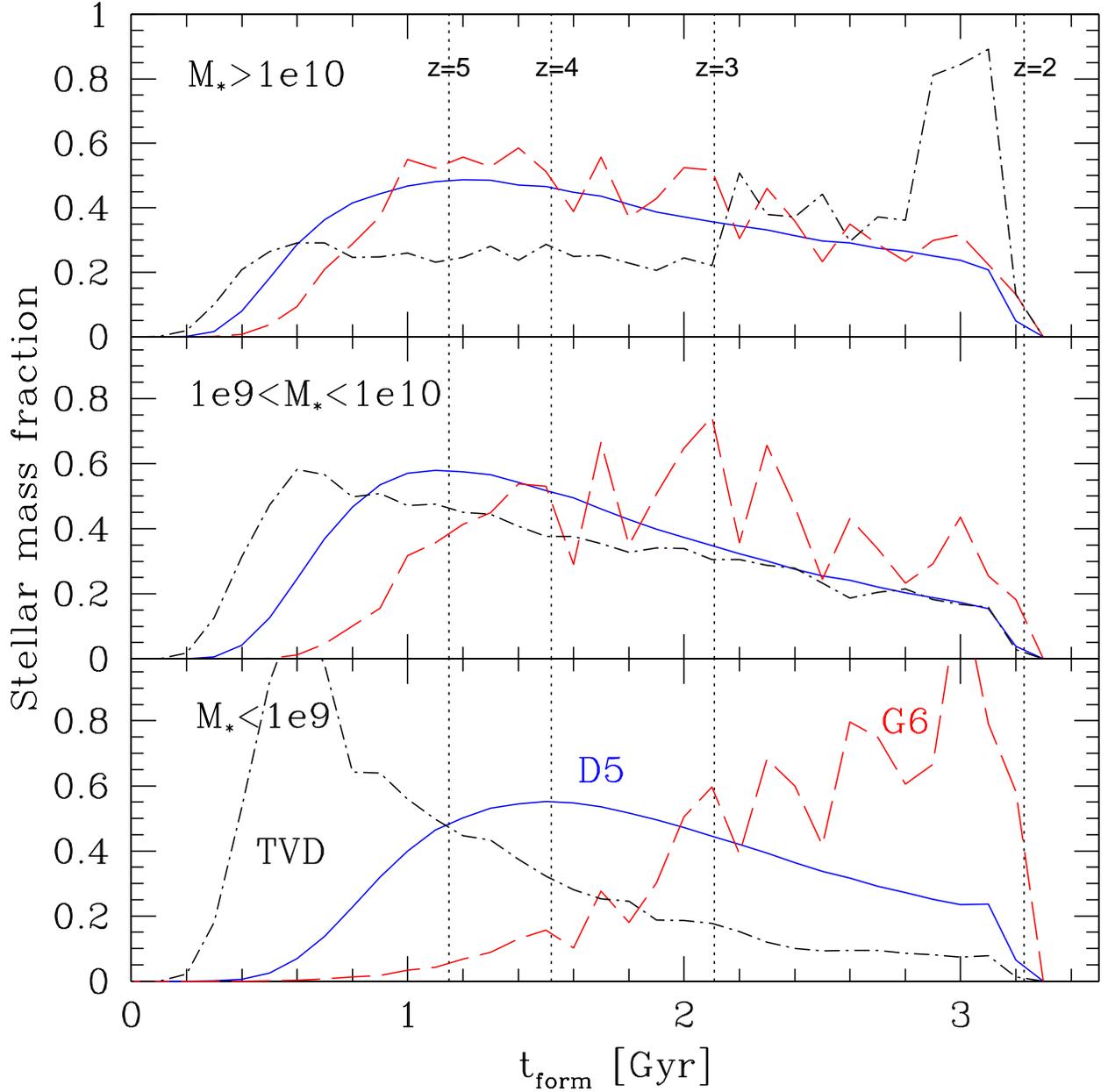}
\caption{Distribution of stellar masses in the simulations as a 
function of formation time ($t_{\rm form}=0$ corresponds to the 
Big Bang). The galaxy sample is divided into 3 categories by their 
stellar mass: $\Mstar>10^{10}\himsun$ (top panel), 
$10^9<\Mstar<10^{10}\himsun$ (middle panel), 
$\Mstar<10^9\himsun$ (bottom panel). 
Different line types correspond to different simulations:
red long-dashed (SPH G6 run), blue solid (SPH D5 run), and 
black dot-dashed (TVD N864L22 run). The vertical dotted lines
indicate four epochs of $z=2$, 3, 4, \& 5, as indicated in the top
panel.
}
\label{age_dist.eps}
\end{figure}

Another way of looking at the star formation history of galaxies is to look
at the distribution of formation times of stellar particles in the
simulation for a given set of galaxies as a whole; i.e. we here analyze the
combined star formation history for a particular class of galaxies 
divided by their stellar mass. 
In Figure~\ref{age_dist.eps}, we show the mass-weighted probability 
distribution of stellar masses as a function of formation time 
($t_{\rm form}=0$ corresponds to the Big Bang),
for different samples of galaxies. To this end, we have split the total
galaxy sample into 3 categories by stellar mass at $z=2$:
$\Mstar>10^{10}\himsun$ (top panel), $10^9<\Mstar<10^{10}\himsun$ (middle
panel), and $\Mstar<10^9\himsun$ (bottom panel).  Different line types
correspond to different simulations: red long-dashed (SPH G6 run), blue
solid (SPH D5 run), and black dot-dashed (TVD N864L22 run). The vertical
dotted lines indicate four different epochs, namely $z=2$, 3, 4, \& 5, as
indicated in the top panel.

In the two top panels for massive galaxies, the distribution roughly agrees
between the different runs, except for the large bump in the TVD run close
to $z=2$, which is caused by starbursts in a few massive galaxies. Also
note that the distribution of G6 run is slightly more bumpy owing to lack
of mass resolution compared to other runs.  

In the bottom panel for the less massive galaxies, the distribution is very
different between the 3 runs. In the TVD run, the bulk of stars in low-mass
galaxies forms before $z=4$, whereas the opposite is true for the SPH G6
run. These low-mass galaxies that formed very early on appeared as a strong 
concentration of points at $G-R\sim 1.2$ in Fig.~\ref{mstar_col.eps} for 
the TVD N864L22 run. 
The result for the SPH D5 appears to be intermediate between the two.
This could perhaps be caused by the additional small-scale power in the TVD
run compared with the SPH runs. The initial mean inter-particle separations
of both dark matter and gas particles in the two SPH runs are comoving 104
and $206~\hikpc$ for the D5 and G6 run, respectively, but the mean
inter-particle separation of dark matter particles in the TVD run is
comoving $51~\hikpc$, with a baryonic cell size of $25~\hikpc$. Therefore,
at very early times, the TVD run should be able to resolve smaller matter
fluctuations than the two SPH runs. Such an effect could explain the
earlier formation epoch of low-mass galaxies in the TVD run compared to the
two SPH runs, provided it is not compensated by the comparatively lower
gravitational force resolution (roughly 2 cells) of the TVD particle-mesh 
method.


\section{Discussion \& Conclusions}
\label{sec:discussion}

We have used two different types of hydrodynamic cosmological simulations
(Eulerian TVD and SPH) to study the properties of massive galaxies at $z=2$
in a $\Lam$CDM universe, with particular emphasis on an observationally
inspired selection based on the $U_n, G, R$ filter set.  The simulated
galaxies at $z=2$ satisfy well the color-selection criteria proposed by
\citet{Ade04} and \citet{Steidel04} when we assume \citet{Calzetti00}
extinction with $E(B-V)=0.15$.  However, we find that the fraction of
stellar mass contained in galaxies that pass the color-selection criteria
could be as low as 50\% of the total.  The number density of simulated
galaxies brighter than $R<25.5$ at $z=2$ is about $2\times 
10^{-2}~h^3~\mpc^{-3}$ for $E(B-V)=0.15$ in the most representative 
run (SPH G6 run), roughly one order of magnitude larger than that of 
Lyman break galaxies at $z=3$. 
The increase of the number density of bright galaxies can be viewed 
as a consequence of the ongoing hierarchical build-up of larger and 
brighter galaxies with ever more stellar mass.

The most massive galaxies at $z=2$ have stellar masses of $\gtsim
10^{11}\Msun$, and their observed-frame $G-R$ colors lie in the range
$0.0<G-R<1.0$ provided that the median extinction is $E(B-V)\sim 0.15$. 
They have been continuously forming stars with a rate
exceeding $30~\Msun~\yr^{-1}$ over a few Gyrs from $z=10$ to $z=2$.
Typically, half of their stellar mass was already assembled by $z\sim 4$.
The bluest galaxies with $-0.2<G-R<0.0$ at $z=2$ are somewhat less massive,
with $\Mstar<10^{11}\himsun$, and are less dominated by old stellar
populations.  Our study suggests that the majority of the most massive
galaxies at $z=2$ could be detected at rest-frame UV wavelengths contrary
to some recent claims based on near-IR studies \citep[e.g.][]{Franx, Dokkum04}
of galaxies at the same epoch, provided the median extinction is in the
range $E(B-V)<0.3$ as indicated by the surveys of Lyman break galaxies at
$z=3$. We plan to extend our analysis to the near-IR pass-bands in
subsequent work.  We find that these massive galaxies reside in the highest
overdensity regions in the simulations at $z=2$.

While it is encouraging to see that the two very different types of
hydrodynamic simulation methods give a similar picture for the overall
properties of massive galaxies at $z=2$, we have also seen some notable
differences in the results of the SPH and TVD simulations throughout the
paper. 

For example, the star formation history of galaxies in the TVD
simulation is clearly more sporadic or `bursty' than that in the SPH
simulations. This probably owes to the differences in resolution and
the details of 
how star formation is modeled in the two codes, as we described in 
Section~\ref{sec:simulation}.  While the dependence of the star formation 
rate on local gas density is essentially the same for the TVD and SPH 
simulations, the different numerical coefficients effectively multiplying 
the local density on the right hand sides of Equations (\ref{eq:TVD}) and 
(\ref{eq:SPH}) produce different temporal smoothing of the local star 
formation rates and the parameters adopted by the TVD code 
leads to a more "bursty" output
($\dot\rho_\star^{\rm TVD} / \dot\rho_\star^{\rm SPH} \simeq 3.8$), 
although not to a significantly different total star formation rate. 
At this point we simply must await a more detailed theory or be guided 
by observations in choosing between the two approaches.

Another difference we saw is that the faint-end slope of the rest-frame 
$V$-band luminosity function: the TVD simulation gives $\alpha=-1.15$, 
similar to what is observed locally $\alpha=-1.2$ \citep[e.g.][]{Blanton01}, 
but the SPH D5 and G6 runs give much steeper slope of $\alpha=-1.8$.
The question of the nature of the faint-end slope of the luminosity 
function is one that we have not resolved at this point.
The difference may owe to the difference in the treatment of 
feedback, but \cite{Chiu01} found a flat faint-end slope of $\alpha=-1.2$ 
in a high-resolution softened Lagrangian hydrodynamic (SLH) simulation 
\citep{Gnedin95} even when the code did not include SN feedback, suggesting
that the photoionization of the gas in low-mass halos by the UV background 
radiation field is responsible for suppressing the formation of low-mass 
galaxies \citep[see also][]{Thoul96, Quinn96, Bullock00}. 
Thus the theoretical expectation is not fully clear at present, and
it will be very interesting to see what is indicated by upcoming
observational programs of deeper high redshift galaxies and future 
theoretical studies.

The present study shows that despite a considerably different numerical
methodology and large differences in the assumed model for star formation,
two independent hydrodynamic codes predict the formation of rather
massive galaxies at $z \simeq 2$ in the $\Lambda$CDM cosmology, with
properties that are broadly consistent with current observations. The
results do not suggest that hierarchical galaxy formation fails to be able
to account for these objects.
On the contrary, it predicts that significant additional mass in 
galaxies will be found as the observational selection criteria
are broadened.


\acknowledgments 

We thank Scott Burles, Hsiao-Wen Chen, Masataka Fukugita, 
Max Pettini, Alice Shapley, and Rob Simcoe for enlightening 
discussions. We also thank Nick Gnedin for refereeing the paper
and for constructive comments on the manuscript.
This work was supported in part by NSF grants ACI 96-19019, 
AST 00-71019, AST 02-06299,
and AST 03-07690, and
NASA ATP grants NAG5-12140, NAG5-13292, and NAG5-13381.
The SPH simulations were performed at the Center for Parallel
Astrophysical Computing at Harvard-Smithsonian Center for
Astrophysics. The TVD simulations were performed at the National
Center for Supercomputing Applications (NCSA).



\begin{deluxetable}{cccccc}  
\tablecolumns{6}  
\tablewidth{0pc}  
\tablecaption{Simulations}
\tablehead{
\colhead{Run} & \colhead{$\Lbox$ [$\himpc$]} & \colhead{$N_{\rm mesh/ptcl}$} & \colhead{$m_{\rm DM}$ [$\himsun$]}  & \colhead{$m_{\rm gas}$ [$\himsun$]} & \colhead{$\Del\ell$ [$\hikpc$]}   
}
\startdata
TVD: N864L22$^a$ & 22.0 & $864^3$ & $8.9\times 10^6$ & $2.2\times 10^5$ & 25.5\cr
SPH: D5$^b$ & 33.75 & $324^3$ & $8.2\times 10^7$ & $1.3\times 10^7$ & 4.2\cr
SPH: G6$^b$ & 100.0 & $486^3$ & $6.3\times 10^8$ & $9.7\times 10^7$ & 5.3\cr
\enddata
\tablecomments{
Parameters of the primary simulations on which this study is based.
The quantities listed are as follows: $\Lbox$ is the simulation 
box size, $N_{\rm mesh/ptcl}$ is the number of the hydrodynamic mesh 
points for TVD or the number of gas particles for SPH, $m_{\rm DM}$ 
is the dark matter particle mass, $m_{\rm gas}$ is the mass of the 
baryonic fluid elements in a grid cell for TVD or the masses of the 
gas particles in the SPH simulations. Note that TVD uses $432^3$ 
dark matter particles for N864 runs. $\Del\ell$ is the size of the 
resolution element (cell size in TVD and gravitational softening 
length in SPH in comoving coordinates; for proper distances, divide 
by $1+z$). The upper indices on the run names correspond to the 
following sets of cosmological parameters: 
$(\Om, \Ol, \Ob, h, n, \sigma_8) = 
(0.29, 0.71, 0.047, 0.7, 1.0, 0.85)$ for (a), and 
$(0.3, 0.7, 0.04, 0.7, 1.0, 0.9)$ for (b).
}
\label{table:simulation}
\end{deluxetable}  


\begin{thebibliography}{102}
\expandafter\ifx\csname natexlab\endcsname\relax\def\natexlab#1{#1}\fi

\bibitem[{Abraham {et~al.}(2004)Abraham, Glazebrook, andD. Crampton,
  Murowinski, J{\o}rgensen, Roth, Hook, Savaglio, Chen, {et~al.}}]{Abraham04}
Abraham, R.~G., Glazebrook, K., andD. Crampton, P. J.~M., Murowinski, R.,
  J{\o}rgensen, I., Roth, K., Hook, I.~M., Savaglio, S., Chen, H.-W., {et~al.}
  2004, AJ, 127, 2455

\bibitem[{Adelberger \& Steidel(2000)}]{Ade00}
Adelberger, K.~L. \& Steidel, C.~C. 2000, ApJ, 544, 218

\bibitem[{Adelberger {et~al.}(2004)Adelberger, Steidel, Shapley, Hunt, Erb,
  Reddy, \& Pettini}]{Ade04}
Adelberger, K.~L., Steidel, C.~C., Shapley, A.~E., Hunt, M.~P., Erb, D.~K.,
  Reddy, N.~A., \& Pettini, M. 2004, ApJ, in press (astro-ph/0401445

\bibitem[{Adelberger {et~al.}(2003)Adelberger, Steidel, Shapley, \&
  Pettini}]{Ade03}
Adelberger, K.~L., Steidel, C.~C., Shapley, A.~E., \& Pettini, M. 2003, ApJ,
  584, 45

\bibitem[{Barton {et~al.}(2004)Barton, Dav\'{e}, Smith, Papovich, Hernquist, \&
  Springel}]{Bart04}
Barton, E.~J., Dav\'{e}, R., Smith, J.-D.~T., Papovich, C., Hernquist, L., \&
  Springel, V. 2004, ApJ, 605, L1

\bibitem[{Beelen {et~al.}(2004)Beelen, Cox, Pety, Carilli, Bertoldi, Momjian,
  Omont, Petitjean, \& Petric}]{Beelen04}
Beelen, A., Cox, P., Pety, J., Carilli, C.~L., Bertoldi, F., Momjian, E.,
  Omont, A., Petitjean, P., \& Petric, A.~O. 2004, A\&A, submitted
  (astro-ph/0404172)

\bibitem[{Blanton {et~al.}(2001)Blanton, Dalcanton, Eisenstein, Loveday,
  Strauss, SubbaRao, Weinberg, Anderson, Annis, Bahcall, {et~al.}}]{Blanton01}
Blanton, M.~R., Dalcanton, J., Eisenstein, D., Loveday, J., Strauss, M.~A.,
  SubbaRao, M., Weinberg, D.~H., Anderson, J. E.~J., Annis, J., Bahcall, N.,
  {et~al.} 2001, AJ, 121, 2358

\bibitem[{Brinchmann \& Ellis(2000)}]{Brinch}
Brinchmann, J. \& Ellis, R. 2000, ApJ, 536, L77

\bibitem[{Bruzual \& Charlot(2003)}]{BClib03}
Bruzual, G. \& Charlot, S. 2003, MNRAS, 344, 1000

\bibitem[{Bullock {et~al.}(2000)Bullock, Kravtsov, \& Weinberg}]{Bullock00}
Bullock, J.~S., Kravtsov, A.~V., \& Weinberg, D.~H. 2000, ApJ, 539, 517

\bibitem[{Calzetti {et~al.}(2000)Calzetti, Armus, Bohlin, Kinney, Koornneef, \&
  Storchi-Bergmann}]{Calzetti00}
Calzetti, D., Armus, L., Bohlin, R.~C., Kinney, A.~L., Koornneef, J., \&
  Storchi-Bergmann, T. 2000, ApJ, 533, 682

\bibitem[{Cen(1992)}]{Cen92}
Cen, R. 1992, ApJS, 78, 341

\bibitem[{Cen {et~al.}(1994)Cen, Miralda-Escude, Ostriker, \& Rauch}]{CO94}
Cen, R., Miralda-Escude, J., Ostriker, J.~P., \& Rauch, M. 1994, ApJ, 437, L9

\bibitem[{Cen \& Ostriker(1992)}]{CO92}
Cen, R. \& Ostriker, J.~P. 1992, ApJ, 399, L113

\bibitem[{Cen \& Ostriker(1993)}]{CO93}
---. 1993, ApJ, 417, 404

\bibitem[{Cen \& Ostriker(1999a)}]{CO99a}
---. 1999a, ApJ, 514, 1

\bibitem[{Cen \& Ostriker(1999b)}]{CO99b}
---. 1999b, ApJ, 519, L109

\bibitem[{Cen \& Ostriker(2000)}]{CO00}
---. 2000, ApJ, 538, 83

\bibitem[{Cen {et~al.}(2003)Cen, Ostriker, Prochaska, \& Wolfe}]{Cen03}
Cen, R., Ostriker, J.~P., Prochaska, J.~X., \& Wolfe, A.~M. 2003, ApJ, 598, 741

\bibitem[{Chabrier(2003a)}]{Chab03a}
Chabrier, G. 2003a, ApJ, 586, L133

\bibitem[{Chabrier(2003b)}]{Chab03b}
---. 2003b, PASP, 115, 763

\bibitem[{Chapman {et~al.}(2003)Chapman, Blain, Ivison, \& Smail}]{Chapman03}
Chapman, S.~C., Blain, A.~W., Ivison, R.~J., \& Smail, I.~R. 2003, Nature, 422,
  695

\bibitem[{Chen \& Marzke(2004)}]{Chen04}
Chen, H.-W. \& Marzke, R. 2004, ApJL, submitted (astro-ph/0405432)

\bibitem[{Chen {et~al.}(2003)Chen, Marzke, McCarthy, Martini, Carlberg,
  Persson, Bunker, Bridge, \& Abraham}]{Chen03}
Chen, H.-W., Marzke, R., McCarthy, P.~J., Martini, P., Carlberg, R.~G.,
  Persson, S.~E., Bunker, A., Bridge, C.~R., \& Abraham, R.~G. 2003, ApJ, 586,
  745

\bibitem[{Chiu {et~al.}(2001)Chiu, Gnedin, \& Ostriker}]{Chiu01}
Chiu, W.~A., Gnedin, N.~Y., \& Ostriker, J.~P. 2001, ApJ, 563, 21

\bibitem[{Cimatti {et~al.}(2003)Cimatti, Daddi, Cassata, Pignatelli, Fasano,
  Vernet, Fomalont, Kellermann, Zamorani, Mignoli, {et~al.}}]{Cimatti03}
Cimatti, A., Daddi, E., Cassata, P., Pignatelli, E., Fasano, G., Vernet, J.,
  Fomalont, E., Kellermann, K., Zamorani, G., Mignoli, M., {et~al.} 2003, A\&A,
  412, L1

\bibitem[{Cohen(2002)}]{Cohen}
Cohen, J.~G. 2002, ApJ, 567, 672

\bibitem[{Cole {et~al.}(2001)Cole, Norberg, Baugh, Frenk, Bland-Hawthorn,
  Bridges, Cannon, Colless, Collins, Couch, {et~al.}}]{Cole}
Cole, S., Norberg, P., Baugh, C.~M., Frenk, C.~S., Bland-Hawthorn, J., Bridges,
  T., Cannon, R., Colless, M., Collins, C., Couch, W., {et~al.} 2001, MNRAS,
  326, 255

\bibitem[{Connolly {et~al.}(1997)Connolly, Szalay, Dickinson, SubbaRao, \&
  Brunner}]{Connolly97}
Connolly, A.~J., Szalay, A.~S., Dickinson, M.~E., SubbaRao, M.~U., \& Brunner,
  R.~J. 1997, ApJ, 486, L11

\bibitem[{Cowie {et~al.}(1999)Cowie, Songaila, \& Barger}]{Cowie99}
Cowie, L.~L., Songaila, A., \& Barger, A.~J. 1999, AJ, 118, 603

\bibitem[{Daddi {et~al.}(2004)Daddi, Cimatti, Renzini, Vernet, Conselice,
  Pozzetti, Mignoli, Tozzi, Broadhurst, de~Serego~Alighieri,
  {et~al.}}]{Daddi04}
Daddi, E., Cimatti, A., Renzini, A., Vernet, J., Conselice, C., Pozzetti, L.,
  Mignoli, M., Tozzi, P., Broadhurst, T., de~Serego~Alighieri, S., {et~al.}
  2004, ApJ, 600, L127

\bibitem[{Dav\'{e} {et~al.}(1999)Dav\'{e}, Hernquist, Katz, \&
  Weinberg}]{Dave99}
Dav\'{e}, R., Hernquist, L., Katz, N., \& Weinberg, D.~H. 1999, ApJ, 511, 521

\bibitem[{Dickinson {et~al.}(2003{\natexlab{a}})Dickinson, Papovich, Ferguson,
  \& Budav\'{a}ri}]{Dick03a}
Dickinson, M., Papovich, C., Ferguson, H., \& Budav\'{a}ri, T.
  2003{\natexlab{a}}, ApJ, 587, 25

\bibitem[{Dickinson {et~al.}(2003{\natexlab{b}})Dickinson, Stern, Giavalisco,
  Ferguson, Tsvetanov, Chornock, Cristiani, Dawson, Dey, Filippenko, Moustakas,
  Nonino, Papovich, Ravindranath, Riess, Rosati, Spinrad, Vanzella,
  {et~al.}}]{Dick03b}
Dickinson, M., Stern, D., Giavalisco, M., Ferguson, H.~C., Tsvetanov, Z.,
  Chornock, R., Cristiani, S., Dawson, S., Dey, A., Filippenko, A.~V.,
  Moustakas, L.~A., Nonino, M., Papovich, C., Ravindranath, S., Riess, A.,
  Rosati, P., Spinrad, H., Vanzella, E., {et~al.} 2003{\natexlab{b}}, ApJ, 600,
  L99

\bibitem[{Eisenstein \& Hut(1998)}]{HOP}
Eisenstein, D.~J. \& Hut, P. 1998, ApJ, 498, 137

\bibitem[{Elston {et~al.}(1988)Elston, Rieke, \& Rieke}]{Elston88}
Elston, R., Rieke, G.~H., \& Rieke, M. 1988, ApJ, 331, L77

\bibitem[{Fan {et~al.}(2001)Fan, Strauss, Schneider, Gunn, Lupton, Becker,
  Davis, Newman, Richards, Gordon, White, {et~al.}}]{Fan01b}
Fan, X., Strauss, M.~A., Schneider, D.~P., Gunn, J.~E., Lupton, R.~H., Becker,
  R.~H., Davis, M., Newman, J.~A., Richards, G.~T., Gordon, R.~L., White,
  R.~L., {et~al.} 2001, AJ, 121, 54

\bibitem[{Fontana {et~al.}(2003)Fontana, Donnarumma, Vanzella, Giallongo,
  Menci, Nonino, Saracco, Cristiani, D'Odorico, Poli, {et~al.}}]{Fontana03}
Fontana, A., Donnarumma, I., Vanzella, E., Giallongo, E., Menci, N., Nonino,
  M., Saracco, P., Cristiani, S., D'Odorico, S., Poli, F., {et~al.} 2003, ApJ,
  594, L9

\bibitem[{Franx {et~al.}(2003)Franx, Labbe, Rudnick, van Dokkum, Daddi,
  F\"{o}rster, Natascha, Moorwood, Rix, R\"{o}ttgering, van~de Wel,
  {et~al.}}]{Franx}
Franx, M., Labbe, I., Rudnick, G., van Dokkum, P.~G., Daddi, E., F\"{o}rster,
  S., Natascha, M., Moorwood, A., Rix, H.-W., R\"{o}ttgering, H., van~de Wel,
  A., {et~al.} 2003, ApJ, 587, L79

\bibitem[{Furlanetto {et~al.}(2004e)Furlanetto, Hernquist, \&
  Zaldarriaga}]{Fetal04e}
Furlanetto, S.~R., Hernquist, L., \& Zaldarriaga, M. 2004e, MNRAS, in press
  (astro-ph/0406131)

\bibitem[{Furlanetto {et~al.}(2003)Furlanetto, Schaye, Springel, \&
  Hernquist}]{Fetal03}
Furlanetto, S.~R., Schaye, J., Springel, V., \& Hernquist, L. 2003, ApJ, 599,
  L1

\bibitem[{Furlanetto {et~al.}(2004d)Furlanetto, Schaye, Springel, \&
  Hernquist}]{Fetal04d}
---. 2004d, ApJ, 606, 221

\bibitem[{Furlanetto {et~al.}(2004a)Furlanetto, Sokasian, \&
  Hernquist}]{Fetal04a}
Furlanetto, S.~R., Sokasian, A., \& Hernquist, L. 2004a, MNRAS, 347, 187

\bibitem[{Furlanetto {et~al.}(2004b)Furlanetto, Zaldarriaga, \&
  Hernquist}]{Fetal04b}
Furlanetto, S.~R., Zaldarriaga, M., \& Hernquist, L. 2004b, ApJ, in press
  (astro-ph/0404112)

\bibitem[{Furlanetto {et~al.}(2004c)Furlanetto, Zaldarriaga, \&
  Hernquist}]{Fetal04c}
---. 2004c, ApJ, in press (astro-ph/0403697)

\bibitem[{Glazebrook {et~al.}(2004)Glazebrook, Abraham, McCarthy, Savaglio,
  Chen, Crampton, Murowinski, Jorgensen, Roth, Hook, Marzke, \&
  Carlberg}]{Glazebrook04}
Glazebrook, K., Abraham, R., McCarthy, P., Savaglio, S., Chen, H.-W., Crampton,
  D., Murowinski, R., Jorgensen, I., Roth, K., Hook, I., Marzke, R., \&
  Carlberg, R. 2004, Nature, in press (astro-ph/0401037)

\bibitem[{Gnedin(1995)}]{Gnedin95}
Gnedin, N.~Y. 1995, ApJS, 97, 231

\bibitem[{Granato {et~al.}(2000)Granato, Lacey, Silva, Bressan, Baugh, Cole, \&
  Frenk}]{Granato00}
Granato, G.~L., Lacey, C.~G., Silva, L., Bressan, A., Baugh, C.~M., Cole, S.,
  \& Frenk, C.~S. 2000, ApJ, 542, 710

\bibitem[{Hernquist(1993)}]{Her93}
Hernquist, L. 1993, ApJ, 404, 717

\bibitem[{Hernquist \& Springel(2003)}]{Her03}
Hernquist, L. \& Springel, V. 2003, MNRAS, 341, 1253

\bibitem[{Hu {et~al.}(1999)Hu, McMahon, \& Cowie}]{Hu99}
Hu, E.~M., McMahon, R.~G., \& Cowie, L.~L. 1999, ApJ, 522, L9

\bibitem[{Hu \& Ridgway(1994)}]{Hu94}
Hu, E.~M. \& Ridgway, S.~E. 1994, AJ, 107, 1303

\bibitem[{Iwata {et~al.}(2003)Iwata, Ohta, Tamura, Ando, Wada, Watanabe,
  Akiyama, \& Aoki}]{Iwata}
Iwata, I., Ohta, K., Tamura, N., Ando, M., Wada, S., Watanabe, C., Akiyama, M.,
  \& Aoki, K. 2003, PASJ, 55, 415

\bibitem[{Katz {et~al.}(1996)Katz, Weinberg, \& Hernquist}]{KWH96}
Katz, N., Weinberg, D.~H., \& Hernquist, L. 1996, ApJS, 105, 19

\bibitem[{Kennicutt(1998)}]{Kennicutt98}
Kennicutt, R. C.~J. 1998, ApJ, 498, 541

\bibitem[{Kodaira {et~al.}(2003)Kodaira, Taniguchi, Kashikawa, Kaifu, Ando,
  Karoji, Ajiki, Akiyama, Aoki, Doi, {et~al.}}]{Kodaira}
Kodaira, K., Taniguchi, Y., Kashikawa, N., Kaifu, N., Ando, H., Karoji, H.,
  Ajiki, M., Akiyama, M., Aoki, K., Doi, M., {et~al.} 2003, PASJ, 55, L17

\bibitem[{Kroupa(2001)}]{Kroupa01}
Kroupa, P. 2001, MNRAS, 322, 231

\bibitem[{Lilly {et~al.}(1996)Lilly, F\`{e}vre, Hammer, \& Crampton}]{Lilly96}
Lilly, S.~J., F\`{e}vre, O.~L., Hammer, F., \& Crampton, D. 1996, ApJ, 460, L1

\bibitem[{Madau(1995)}]{Madau95}
Madau, P. 1995, ApJ, 441, 18

\bibitem[{McCarthy {et~al.}(1992)McCarthy, Persson, \& West}]{McCarthy92}
McCarthy, P.~J., Persson, S.~E., \& West, S.~C. 1992, ApJ, 386, 52

\bibitem[{Menci {et~al.}(2002)Menci, Cavaliere, Fontana, Giallongo, \&
  Poli}]{Menci02}
Menci, N., Cavaliere, A., Fontana, A., Giallongo, E., \& Poli, F. 2002, ApJ,
  575, 18

\bibitem[{Nagamine(2002)}]{Nag02}
Nagamine, K. 2002, ApJ, 564, 73

\bibitem[{Nagamine {et~al.}(2004{\natexlab{a}})Nagamine, Cen, Hernquist,
  Ostriker, \& Springel}]{Nachos1}
Nagamine, K., Cen, R., Hernquist, L., Ostriker, J.~P., \& Springel, V.
  2004{\natexlab{a}}, ApJ, 610, 45

\bibitem[{Nagamine {et~al.}(2001a)Nagamine, Fukugita, Cen, \&
  Ostriker}]{Nag01a}
Nagamine, K., Fukugita, M., Cen, R., \& Ostriker, J.~P. 2001a, ApJ, 558, 497

\bibitem[{Nagamine {et~al.}(2001b)Nagamine, Fukugita, Cen, \&
  Ostriker}]{Nag01b}
---. 2001b, MNRAS, 327, L10

\bibitem[{Nagamine {et~al.}(2004a)Nagamine, Springel, \& Hernquist}]{NSH04a}
Nagamine, K., Springel, V., \& Hernquist, L. 2004a, MNRAS, 348, 421

\bibitem[{Nagamine {et~al.}(2004b)Nagamine, Springel, \& Hernquist}]{NSH04b}
---. 2004b, MNRAS, 348, 435

\bibitem[{Nagamine {et~al.}(2004{\natexlab{b}})Nagamine, Springel, Hernquist,
  \& Machacek}]{NSHM}
Nagamine, K., Springel, V., Hernquist, L., \& Machacek, M. 2004{\natexlab{b}},
  MNRAS, 350, 385

\bibitem[{Neyrinck {et~al.}(2004)Neyrinck, Gnedin, \& Hamilton}]{Neyrinck04}
Neyrinck, M.~C., Gnedin, N.~Y., \& Hamilton, A. J.~S. 2004, MNRAS, accepted
  (astro-ph/0402346)

\bibitem[{Omont {et~al.}(2003)Omont, Beelen, Bertoldi, \& others~et
  al.}]{Omont03}
Omont, A., Beelen, A., Bertoldi, F., \& others~et al. 2003, A\&A, 398, 857

\bibitem[{Ouchi {et~al.}(2003a)Ouchi, Shimasaku, Furusawa, Miyazaki, Doi,
  Hamabe, Hayashino, Kimura, Kodaira, Komiyama, {et~al.}}]{Ouchi03a}
Ouchi, M., Shimasaku, K., Furusawa, H., Miyazaki, M., Doi, M., Hamabe, M.,
  Hayashino, T., Kimura, M., Kodaira, K., Komiyama, Y., {et~al.} 2003a, ApJ,
  582, 60

\bibitem[{Ouchi {et~al.}(2003b)Ouchi, Shimasaku, Furusawa, Miyazaki, Doi,
  Hamabe, Hayashino, Kimura, Kodaira, Komiyama, {et~al.}}]{Ouchi03b}
---. 2003b, ApJ, submitted (astro-ph/0309655)

\bibitem[{Pascarelle {et~al.}(1998)Pascarelle, Lanzetta, \&
  Fernandez-Soto}]{Pas98}
Pascarelle, S.~M., Lanzetta, K.~M., \& Fernandez-Soto, A. 1998, ApJ, 508, L1

\bibitem[{Pei(1995)}]{Pei95}
Pei, Y.~C. 1995, ApJ, 438, 623

\bibitem[{Poli {et~al.}(2003)Poli, Giallongo, Fontana, Menci, Zamorani, Nonino,
  Saracco, Vanzella, Donnarumma, Salimbeni, {et~al.}}]{Poli}
Poli, F., Giallongo, E., Fontana, A., Menci, N., Zamorani, G., Nonino, M.,
  Saracco, P., Vanzella, E., Donnarumma, I., Salimbeni, S., {et~al.} 2003, ApJ,
  593, L1

\bibitem[{Quinn {et~al.}(1996)Quinn, Katz, \& Efstathiou}]{Quinn96}
Quinn, T., Katz, N., \& Efstathiou, G. 1996, MNRAS, 278, 49

\bibitem[{Rhoads \& Malhotra(2001)}]{Rhoads}
Rhoads, J.~E. \& Malhotra, S. 2001, ApJ, 563, L5

\bibitem[{Robertson {et~al.}(2004)Robertson, Yoshida, Springel, \&
  Hernquist}]{Robertson04}
Robertson, B.~E., Yoshida, N., Springel, V., \& Hernquist, L. 2004, ApJ, 606,
  32

\bibitem[{Rudnick {et~al.}(2003)Rudnick, Rix, Franx, Labbe, Blanton, Daddi,
  F\"{o}rster, Natascha, Moorwood, R\"{o}ttgering, Trujillo,
  {et~al.}}]{Rudnick03}
Rudnick, G., Rix, H.-W., Franx, M., Labbe, I., Blanton, M., Daddi, E.,
  F\"{o}rster, S., Natascha, M., Moorwood, A., R\"{o}ttgering, H., Trujillo,
  I., {et~al.} 2003, ApJ, 599, 847

\bibitem[{Ryu {et~al.}(1993)Ryu, Ostriker, Kang, \& Cen}]{Ryu93}
Ryu, D., Ostriker, J.~P., Kang, H., \& Cen, R. 1993, ApJ, 414, 1

\bibitem[{Sawicki {et~al.}(1997)Sawicki, Lin, \& Yee}]{Sawicki97}
Sawicki, M.~J., Lin, H., \& Yee, H. K.~C. 1997, AJ, 113, 1

\bibitem[{Schmidt(1968)}]{Schmidt68}
Schmidt, M. 1968, ApJ, 151, 393

\bibitem[{Schmidt {et~al.}(1995)Schmidt, Schneider, \& Gunn}]{Schmidt95}
Schmidt, M., Schneider, D.~P., \& Gunn, J.~E. 1995, AJ, 110, 68

\bibitem[{Shapley {et~al.}(2001)Shapley, Steidel, Adelberger, Dickinson,
  Giavalisco, \& Pettini}]{Shapley01}
Shapley, A.~E., Steidel, C.~C., Adelberger, K.~L., Dickinson, M., Giavalisco,
  M., \& Pettini, M. 2001, ApJ, 562, 95

\bibitem[{Smail {et~al.}(1997)Smail, Ivision, \& Blain}]{Smail97}
Smail, I., Ivision, R.~J., \& Blain, A.~W. 1997, ApJ, 490, L5

\bibitem[{Smail {et~al.}(2002)Smail, Owen, Morrison, Keel, Ivison, \&
  Ledlow}]{Smail02}
Smail, I., Owen, F.~N., Morrison, G.~E., Keel, W.~C., Ivison, R.~J., \& Ledlow,
  M.~J. 2002, ApJ, 581, 844

\bibitem[{Somerville {et~al.}(2001)Somerville, Primack, \& Faber}]{Som}
Somerville, R.~S., Primack, J.~R., \& Faber, S.~M. 2001, MNRAS, 320, 504

\bibitem[{Spergel {et~al.}(2003)Spergel, Verde, Peiris, Komatsu, Nolta,
  Bennett, Halpern, Hinshaw, Jarosik, Kogut, {et~al.}}]{Spergel03}
Spergel, D., Verde, L., Peiris, H.~V., Komatsu, E., Nolta, M.~R., Bennett,
  C.~L., Halpern, M., Hinshaw, G., Jarosik, N., Kogut, A., {et~al.} 2003, ApJS,
  148, 175

\bibitem[{Springel \& Hernquist(2002)}]{SH02}
Springel, V. \& Hernquist, L. 2002, MNRAS, 333, 649

\bibitem[{Springel \& Hernquist(2003{\natexlab{a}})}]{SH03a}
---. 2003{\natexlab{a}}, MNRAS, 339, 289

\bibitem[{Springel \& Hernquist(2003{\natexlab{b}})}]{SH03b}
---. 2003{\natexlab{b}}, MNRAS, 339, 312

\bibitem[{Springel {et~al.}(2001{\natexlab{a}})Springel, Tormen, Kauffmann,
  Yoshida, \& White}]{Springel01}
Springel, V., Tormen, G., Kauffmann, G., Yoshida, N., \& White, S. D.~M.
  2001{\natexlab{a}}, MNRAS, 328, 726

\bibitem[{Springel {et~al.}(2001{\natexlab{b}})Springel, Yoshida, \&
  White}]{Gadget}
Springel, V., Yoshida, N., \& White, S. D.~M. 2001{\natexlab{b}}, New
  Astronomy, 6, 79

\bibitem[{Steidel {et~al.}(2003)Steidel, Adelberger, Adelberger, Shapley,
  Pettini, Dickinson, \& Giavalisco}]{Steidel03}
Steidel, C.~C., Adelberger, K.~L., Adelberger, K.~L., Shapley, A.~E., Pettini,
  M., Dickinson, M., \& Giavalisco, M. 2003, ApJ, 592, 728

\bibitem[{Steidel {et~al.}(1999)Steidel, Adelberger, Giavalisco, Dickinson, \&
  Pettini}]{Steidel99}
Steidel, C.~C., Adelberger, K.~L., Giavalisco, M., Dickinson, M., \& Pettini,
  M. 1999, ApJ, 519, 1

\bibitem[{Steidel \& Hamilton(1993)}]{Steidel93}
Steidel, C.~C. \& Hamilton, D. 1993, AJ, 105, 2017

\bibitem[{Steidel {et~al.}(2004)Steidel, Shapley, Pettini, Adelberger, Erb,
  Reddy, \& Hunt}]{Steidel04}
Steidel, C.~C., Shapley, A.~E., Pettini, M., Adelberger, K.~L., Erb, D.~K.,
  Reddy, N.~A., \& Hunt, M.~P. 2004, ApJ, 604, 534

\bibitem[{Taniguchi {et~al.}(2003)Taniguchi, Ajiki, Murayama, Nagao, Veilleux,
  Sanders, Komiyama, Shioya, Fujita, Kakazu, {et~al.}}]{Taniguchi}
Taniguchi, Y., Ajiki, M., Murayama, T., Nagao, T., Veilleux, S., Sanders,
  D.~B., Komiyama, Y., Shioya, Y., Fujita, S., Kakazu, Y., {et~al.} 2003, ApJ,
  585, L97

\bibitem[{Thoul \& Weinberg(1996)}]{Thoul96}
Thoul, A.~A. \& Weinberg, D.~H. 1996, ApJ, 465, 608

\bibitem[{Treyer {et~al.}(1998)Treyer, Ellis, Millard, Donas, \&
  Bridges}]{Treyer98}
Treyer, M.~A., Ellis, R.~S., Millard, B., Donas, J., \& Bridges, T.~J. 1998,
  MNRAS, 300, 303

\bibitem[{van Dokkum {et~al.}(2004)van Dokkum, Franx, Schreiber, Illingworth,
  Daddi, Knudsen, Labbe, Moorwood, Rix, Rottgering, Rudnick, Trujillo, van~der
  Werf, van~der Wel, van Starkenburg, \& Wuyts}]{Dokkum04}
van Dokkum, P., Franx, M., Schreiber, N.~F., Illingworth, G., Daddi, E.,
  Knudsen, K.~K., Labbe, I., Moorwood, A., Rix, H.-W., Rottgering, H., Rudnick,
  G., Trujillo, I., van~der Werf, P., van~der Wel, A., van Starkenburg, L., \&
  Wuyts, S. 2004, ApJ, in press (astro-ph/0404471)

\bibitem[{Zaldarriaga {et~al.}(2004)Zaldarriaga, Furlanetto, \&
  Hernquist}]{Zetal04}
Zaldarriaga, M., Furlanetto, S.~R., \& Hernquist, L. 2004, ApJ, in press
  (astro-ph/0311514)

\end{thebibliography}
\end{document}